\documentclass[a4paper,11pt]{article}
\pdfoutput=1 

\usepackage{jheppub} 

\usepackage[T1]{fontenc} 
\usepackage{physics}
\NewDocumentCommand{\tens}{t_}
{%
	\IfBooleanTF{#1}
	{\tensop}
	{\otimes}%
}
\NewDocumentCommand{\tensop}{m}
{%
	\mathbin{\mathop{\otimes}\displaylimits_{#1}}%
}
\usepackage{graphicx}
\usepackage{wrapfig}
\usepackage{amsmath}
\usepackage{amsfonts}
\usepackage{amssymb}
\usepackage{quotes}
\usepackage{subcaption}
\usepackage{calligra}

\usepackage{changepage} 

\usepackage{afterpage}

\usepackage{placeins}

\title{\textbf{\boldmath Mixed state entanglement measures for open string geometry}}

\author[1]{Souvik Paul,\note{Corresponding author.}}
\author{Anirban Roy Chowdhury,}
\author{Sunandan Gangopadhyay}

\affiliation[a]{\textit{Department of Astrophysics and High Energy Physics\\}
	\textit{ S.N.~Bose National Centre for Basic Sciences,\\}
	\textit{JD Block, Sector-III, Salt Lake, Kolkata 700106, India}}

\emailAdd{souvik.paul@bose.res.in}
\emailAdd{iamanirban@bose.res.in}
\emailAdd{sunandan.gangopadhyay@bose.res.in}

    \abstract{\noindent In the gauge/gravity framework flavor degrees of freedom are introduced to the gauge theory side through the insertion of probe flavor $D$ branes in the supergravity background. Scalar, vector or spinor fluctuations on theses flavor branes perceive neither the background supergravity metric nor the worldvolume-induced metric of the flavor branes. Indeed, they follow an effective metric called the open string metric (OSM). Through a proper choice of worldvolume gauge fields, one can induce a horizon structure in these OSMs. Studying holographic information-theoretic quantities in these kinds of geometries is equivalent to studying information-theoretic quantities in the flavor sector on the gauge theory side. In this article, we have studied various information-theoretic measures like mutual information, entanglement wedge cross-section and entanglement negativity for an open string geometry. All the calculations are done by considering strip-like parallel boundary subsystems for three- and four-dimensional OSMs. We have also measured the change in these mixed-state information-theoretic quantities from the relevant quantities in the pure AdS geometry due to the application of a background electric field.}
\begin{document}
\maketitle
	\flushbottom
\section{Introduction}
In the context of string theory, there are two different kinds of strings that is open and closed strings \cite{green2012superstring1,green2012superstring2,polchinski1999string,zwiebach2004first,lust1989lectures}. In four dimensions, the open string oscillations have two degrees of freedom. These degrees of freedom represent two polarizations of a gauge field. In this sense, the open strings represent a gauge theory. On the other hand, the closed string oscillations occur in two simultaneous directions. This property of the closed string oscillations explains the spin-$2$ nature of the graviton. In this sense, the closed strings represent a theory of gravity. Two open strings join into one open string. But if the endpoints of two open strings can join, the endpoints of a single open string can join as well. Otherwise, one has to require a nonlocal
constraint on the string dynamics. Thus, the endpoints of a single open string can
join to produce a closed string. Namely, if there are open strings, one has to have
closed strings. This is the reason why we have two kinds of strings. Therefore, this open/closed string duality indicates a duality between a strongly coupled gauge theory and a weakly coupled gravity theory. Based on this phenomenon, Maldacena, in his seminal paper \cite{Maldacena:1997re} had shown that the large $N$ limit of a $\mathcal{N}=4$ supersymmetric Yang-Mills theory (with $SU(N)$ gauge group) is dual to a type IIB superstring theory in AdS$_5\cross$S$^5$ space. This was the stepping stone towards the well-known AdS/CFT correspondence \cite{Maldacena:1997re,Gubser:1998bc,Witten:1998qj,Aharony:1999ti,Natsuume:2014sfa,Nastase:2007kj}, which states that a strongly coupled gauge theory in $d$-dimensions is dual to a weakly coupled gravity theory in $d+1$-dimensions. Over the past few decades this duality has gained attention in different branches of physics starting from condensed matter physics \cite{Sachdev:2010ch,Keski-Vakkuri:2008ffv,Hartnoll:2008vx,Lee:2008xf,Cai:2011ky,Gangopadhyay:2012am,Koutsoumbas:2009pa,Bergman:2010gm,Gangopadhyay:2012gx,Paul:2025apr}, quantum chromodynamics \cite{Csaki:2008dt,Brodsky:2008pf,Erlich:2005qh,Karch:2006pv,Kruczenski:2004me,Aharony:2002up}, quantum gravity \cite{Bak:2006nh,Rovelli:1997na,Engelhardt:2015gla}, cosmology \cite{Nojiri:2000kq,Antonini:2019qkt,Craps:2007ch,Paul:2025gpk,McFadden:2009fg,Banks:2001px,Brax:2003fv}. The original AdS/CFT framework proposed by Maldacena \cite{Maldacena:1997re} only describes gluons in the gauge theory side through a weakly interacting gravity theory with a negative cosmological constant. These gluons transform under the adjoint representation of the $SU(N)$ gauge group. On the other hand, Quantum Chromodynamics (QCD) is a very successful quantum field
theory description of the strong interactions and is by now very well tested experimentally. The theory of QCD consists of quarks that transform in the fundamental representation of a non-abelian $SU(3)$ gauge group. Here, the interactions are mediated by gauge bosons, the gluon fields, which transform in the adjoint of $SU(3)$. An obvious question is whether the AdS/CFT duality can be used to understand the properties of strongly interacting QCD. In order to obtain a QCD-like theory, one needs to break supersymmetry and remove conformal invariance. This is done to achieve a running coupling, as well as to introduce quark fields. On the gravity side, we need extra ingredients to break supersymmetry and introduce the quark degrees of freedom. This can be done by introducing something called the flavor $D$-branes \cite{Karch:2002sh,Karch:2007pd,DeWolfe:2001pq}. For example, in the original setup of Maldacena, the flavor degrees of freedom are added on the gauge theory side by adding flavor $Dp$ branes in the supergravity side, which is generated by a stack of $D3$ or some other brane configuration. From the concept of intersecting $Dp-D(p+4)$ branes \cite{Polchinski:1995mt,Berkooz:1996km,Gauntlett:1997cv,Lust:2004ks,Aldazabal:2000cn}, it is well known that these systems preserve $\frac{1}{4}$ of the bulk supersymmetry. Therefore, these kinds of intersecting branes are BPS $\frac{1}{4}$, and they form a stable configuration. Hence, to introduce flavor degrees of freedom in the boundary field theory, we need to introduce flavor $D7$ branes in the supergravity background sourced by a stack of $D3$ branes. From the AdS/CFT perspective, a stack of $N$ number of $D3$ branes gives rise to a gauge theory in the worldvolume. The world-volume theory is simply a $(3+1)$-dimensional SU($N$) gauge theory in the low-energy limit \cite{Maldacena:1997re}. Since a large number of $D3$ branes are necessary to keep the dual supergravity background weakly coupled, $N$ must be a large number. Open strings, which have charged endpoints, may end on this stack of $D3$ branes. This refers to a world-volume gauge theory with only adjoint degrees of freedom. In the probe limit, when $N_f$ number of flavor $D7$ branes are added in the supergravity background introduces new strings stretched between $D3$ and $D7$ branes and hence generate matter in the fundamental representation \cite{Karch:2002sh,Erdmenger:2007cm,Kruczenski:2003be,Babington:2003vm}. These strings carry a single charge under the SU($N$) group on the $D3$ branes and consequently correspond to quark fields.
\\
In recent years, this AdS/CFT duality has been widely used to study various information-theoretic quantities for strongly interacting CFTs in a holographic manner. Ryu and Takayanagi in \cite{Ryu:2006bv,Ryu:2006ef} first proposed a systematic approach to compute quantum entanglement in strongly interacting boundary CFTs holographically. A huge number of studies have been done in this direction to study the entanglement entropy and different mixed state information theoretic quantities of various strongly coupled boundary field theories. In \cite{Banerjee:2020gyv} the authors have studied the holographic entanglement entropy and holographic volume complexity for the Open String Metric background. It is already argued that in the AdS/CFT picture, studying the OSM in bulk indeed is dual to studying boundary field theory with flavor degrees of freedom, that is, quarks \cite{Banerjee:2016qeu,Banerjee:2020gyv}. Currently, the existing literature \cite{Kontoudi:2013rla,Karch:2014ufa,Kol:2014nqa,Georgiou:2015pia,Jones:2015twa,Vaganov:2015vpq,Kumar:2017vjv,Jokela:2024cxb,Chang:2013mca} offers limited exploration of mixed-state information-theoretic quantities within the flavor sector from a holographic perspective. In this paper, we have calculated various mixed-state information-theoretic quantities like mutual information, entanglement wedge cross section and entanglement negativity for open string metric in $(2+1)$ and ($3+1$)-dimensions. Hence, before moving further, we will discuss various information-theoretic measures from a quantum information-theoretic perspective. \\
The concept of entanglement entropy is quite important to quantify entanglement between two parts of a pure bipartite state. Now, for a pure bipartite state, the entanglement entropy is nothing but the von Neumann entropy of the reduced density matrix of any of the systems. In order to compute the von Neumann entropy, one needs to follow the following procedure. Let us start with a pure bipartite state consisting of two parts $A$ and $B$. The Hilbert space of this kind of system has the form $\mathcal{H}\equiv \mathcal{H}_A\otimes \mathcal{H}_B$, where $A\cup B$ is the total system. The entanglement entropy of system $A$ can be found by computing the von Neumann entropy of the reduced density matrix of $A$. The reduced density matrix of $A$ can be found by tracing out the degrees of freedom of $B$. Therefore, the von Neumann entropy of $A$ is given by \cite{von2013mathematische,nielsen2010quantum}
\begin{equation}
    S_{EE}(A)=-\Tr \rho_A\log \rho_A
\end{equation}
where $\rho_A$ is the reduced density matrix of the subsystem $A$. This reduced density matrix can be computed by taking the partial trace over the degrees of freedom of $B$ on the total density matrix $\rho_{AB}=\ket{\psi}\bra{\psi};~\ket{\psi}\in \mathcal{H}$. Hence the reduced density matrix corresponding to subsystem $A$ is given by $\rho_A=\Tr_{B}\rho_{AB}$. Although the von Neumann entropy is a good measure for quantum entanglement in pure states, for mixed states, it contains both classical and quantum correlations. Thus, it is not a good measure of quantum entanglement for mixed states. Another problem with von Neumann entropy is that it also contains an ultraviolet (UV) divergent term, which appears from short-distance entanglement.\\
With the above mentioned defination of von Neumann entropy, one can also define another useful quantity in the context of quantum information theory, known as mutual information. The mutual information between two systems $A$ and $B$ is defined as \cite{nielsen2010quantum}
\begin{equation}
    I(A:B)=S_{EE}(A)+S_{EE}(B)-S_{EE}(A\cup B)~
\end{equation}
where $S_{EE}(A),~S_{EE}(B)$, and $S_{EE}(A\cup B)$ are the von Neumann entropies of $A$, $B$ and $A\cup B$ respectively. Several interesting studies on mutual information in the context of the black hole information paradox can be found in \cite{Hawking:1975vcx,Hawking:1976ra,Penington:2019kki,Penington:2019npb,Almheiri:2019hni,Hartman:2020swn,Wang:2021mqq,Almheiri:2019psy,Ling:2020laa,Alishahiha:2020qza,Saha:2021ohr,RoyChowdhury:2022awr,RoyChowdhury:2023eol,RoyChowdhury:2026ubx,Cheng:2025bnj,Yu:2025euq}.
It should be mentioned that the von Neumann entropy is free from any UV divergent term. In \cite{Wolf:2007tdq}, it was argued that the mutual information is a measure of the collective correlations between two subsystems. It also contains both classical and quantum correlations. Various correlation measures for mixed states have been proposed in the existing literature, with the entanglement of purification (EoP) standing out as one of the most promising candidates. Purification is basically creating a pure state $\ket{\psi}$ from a mixed state with density matrix ($\rho_{AB}$) by introducing auxiliary degrees of freedom to the original Hilbert space. If $\rho_{AB}$ is the density matrix of a bipartite state then one can introduce auxiliary degrees of freedom $A^{\prime}$ and $B^{\prime}$ with the total Hilbert space $\mathcal{H}$ and construct a pure state such that $\ket{\psi}\in \mathcal{H}_{AA^{\prime}B B^{\prime}}=\mathcal{H}_{A A^{\prime}}\otimes \mathcal{H}_{B B^{\prime}}$ and $\rho_{AB}=\Tr_{A^{\prime}B^{\prime}}\ket{\psi}\bra{\psi}$. Then the entanglement of purification (EoP) for $\rho_{AB}$ is defined as \cite{Terhal:2002riz}
\begin{equation}
    E_P(\rho)=E_P(A,B)=\min_{\ket{\psi}} S_{AA^{\prime}}~
\end{equation}
where the minimization is over all possible purifications of $\rho_{AB}$ and $S_{AA^{\prime}}$ is the von Neumann entropy of the density matrix $\rho_{AA^{\prime}}=\Tr_{BB^{\prime}}\ket{\psi}\bra{\psi}$. If $\rho_{AB}$ is a pure state density matrix, no purification is required and $E_P=S(A)=S(B)$. It should be worth mentioning that EoP is bounded by half of the mutual information, that is
\begin{equation}
    E_P(A,B)\geq \frac{I(A:B)}{2}~.
\end{equation}
Another important mixed-state information-theoretic quantity present in the literature is the entanglement of negativity (EN). This quantity basically sets a maximum limit on the
amount of entanglement that can be distilled from a quantum system in a mixed state. In order to compute the entanglement negativity, one can consider a tripartite system consisting of $A=A_1\cup A_2$ and $B=A^c$. The Hilbert space of the bipartite state $A$ can be written as $\mathcal{H}_A\equiv \mathcal{H}_{A_1}\otimes \mathcal{H}_{A_2}$, where $\mathcal{H}_{A_1}$ and $\mathcal{H}_{A_2}$ are Hilbert spaces of $A_1$ and $A_2$ respectively. The density matrix of the system $A$ is given by
\begin{equation}
    \rho_A=\Tr_{B=A^c} \rho_{AB}~.
\end{equation}
The computation of the entanglement negativity involves taking the partial transpose of the above density matrix. Let us assume $\ket{e_i^{(1)}}$ and $\ket{e_i^{(2)}}$ are the basis vectors corresponding to $A_1$ and $A_2$, then the partial transpose of $\rho_A$ is defined as 
\begin{equation}
    \bra{e_i^{(1)}e_j^{(2)}}\rho_A^{T_2}\ket{e_k^{(1)}e_l^{(2)}}=\bra{e_i^{(1)}e_l^{(2)}}\rho_A\ket{e_k^{(1)}e_j^{(2)}}
\end{equation}
where $\rho_A^{T_2}$ represents the partial transpose of the total density matrix $\rho$ with respect to $A^c$. Entanglement negativity measures the extent to which $\rho_A^{T_2}$ is not positive \cite{Horodecki:1998kf}; this eventually signifies the term `negativity'. In the literature, there are actually two quantities, one is negativity, and the other is the entanglement negativity or logarithmic negativity. The negativity between the subsystems $A_1$ and $A_2$ is basically defined as \cite{Vidal:2002zz} 
\begin{equation}
    \mathcal{N}(\rho)=\frac{||\rho_A^{T_2}||_1-1}{2}~
\end{equation}
where $||\rho_A^{T_2}||$ is the trace norm of $\rho_A^{T_2}$. The above quantity corresponds to the absolute value of the sum of the negative eigenvalues of $\rho_A^{T_2}$. One can also show that $\mathcal{N}(\rho)$ does not increase under local operations and classical communication (LOCC) \cite{Vidal:1998re}. Thus, it is an entanglement monotone. Another quantity is the entanglement negativity or logarithmic negativity. For two subsystems $A_1$ and $A_2$, the logarithmic negativity is defined as follows \cite{Vidal:2002zz}
\begin{equation}
    E_N(\rho)=\ln (||\rho_A^{T_2}||_1)~.
\end{equation}
The above quantity again exhibits some form of monotonicity under LOCC and is an additive quantity. In the context of quantum field theories, it is very challenging to compute these quantities. Nevertheless, gauge/gravity duality offers a systematic approach to compute it through the bulk/boundary correspondence.\\
In this paper, we holographically compute all of these quantities discussed above for open string metrics (OSM) in $(2+1)$ and $(3+1)$-dimensions. This kind of geometry was first encountered in the noncommutative string backgrounds. Here they appear while studying the fluctuation modes on the flavor $D$ branes. Due to the presence of world volume gauge fields, a horizon structure is also induced in the open string geometries. For this kind of specially engineered horizons, the dual boundary state is nothing but a non-equilibrium steady state (NESS). In this sense, studying  OSMs as the bulk theory is equivalent to studying the physics of flavor sector on the gauge theory side. Studying the mixed state information-theoretic quantities for OSMs in $(2+1)$ and $(3+1)$-dimensions is indeed equivalent to studying the mixed state information-theoretic measures in  
In this paper, we holographically compute all of these quantities discussed above for open string metrics (OSM) in $(2+1)$ and $(3+1)$-dimensions. All of our calculations are done while considering strip-like boundary subsystems. This kind of geometry was first encountered in the noncommutative string backgrounds \cite{Seiberg:1999vs}. Here they appear while studying the fluctuation modes on the flavor $D$ branes. Due to the presence of world volume gauge fields, a horizon structure is also induced in the open string geometries. For this kind of specially engineered horizons, the dual boundary state is nothing but a non-equilibrium steady state (NESS) \cite{Kundu:2013eba,Kundu:2015qda,Banerjee:2015cvy}. In this sense, studying  OSMs as the bulk theory is equivalent to studying the physics of flavor sector on the gauge theory side. Studying the mixed state information-theoretic quantities for OSMs in $(2+1)$ and $(3+1)$-dimensions is indeed equivalent to studying the mixed state information-theoretic measures in the flavor sectors of $(1+1)$ and $(2+1)$-dimensional boundary field theory, respectively.\\
This paper is organized in the following way. In section \eqref{Sec:2}, we have given a quick review of the open string geometry. It is shown that while studying the fluctuations on the flavor $D$ branes, the open string geometries emerge. In a supergravity background with an AdS$_3$ metric, a proper choice of flavor brane world volume gauge field gives rise to a horizon structure in the OSM. For a three-dimensional scenario, we have briefly re-derived the OSM. In section \eqref{Sec:3}, we have computed the holographic entanglement entropy (HEE) for $(2+1)$ and $(3+1)$-dimensional OSMs and found the change in HEE from the pure AdS case. In section \eqref{Sec:4}, for $(2+1)$ and $(3+1)$-dimensions, we have derived the expressions for holographic mutual information (HMI) and entanglement wedge cross section (EWCS). We have also graphically represented their variation for different values of the applied electric field. Subsequently, in section \eqref{Sec:5}, we have evaluated the entanglement negativity for the same open string geometries, considering parallel strip-like boundary subsystems in both adjacent and disjoint configurations. Finally, in section \eqref{Sec:6}, we summarize our findings and conclude with some discussions along with possible future directions. We also add an Appendix for the sake of completeness.
\section{Open string geometry}\label{Sec:2}
\begin{figure}[t]
    \centering
    \includegraphics[width=0.9\linewidth]{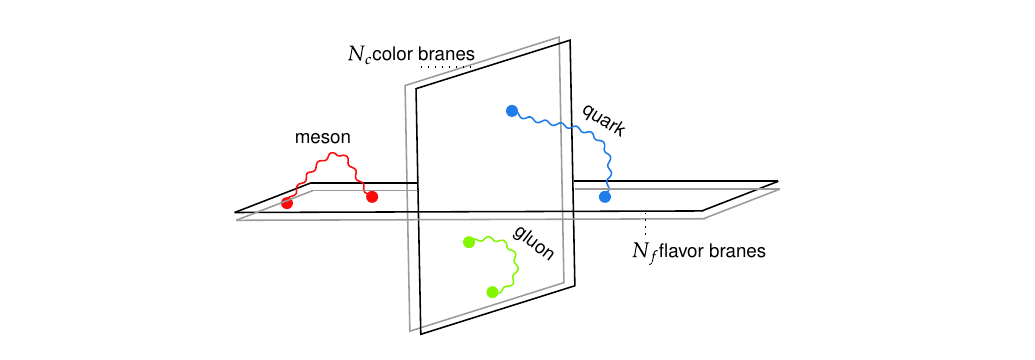}
    \caption{A schematic diagram for intersecting color and flavor branes. Open strings with both endpoints on color branes, one point at flavor and the other at color brane and both ends on flavor branes represent gluons, quarks and mesons, respectively. }
    \label{fig:intersecting branes}
\end{figure}

In this section, we will briefly discuss about open string geometries and the emergence of open string horizons in the presence of world volume gauge fields. Let us start with the action of a flavor $Dp$ brane, which is inserted in the supergravity background (sourced by color branes). The dynamics of these flavor branes are governed by the Dirac-Born-Infeld (DBI) action, which reads
\begin{equation}
    S_{D_p}=\tau_p \int_{\mathcal{M}_{p+1}}d^{p+1}y e^{-\Phi}\sqrt{-\det\Big[P(g_{ab}+B_{ab})+2\pi\alpha^{\prime}F_{ab}\Big]}
\end{equation}
where $\left\{\Phi,g,B\right\}$ are supergravity data in the string frame, which consist of the dilaton, the metric and the NS-NS 2-form, respectively. $P(g_{ab}+B_{ab})$ is the pull back of the background metric, $\tau_p=(2\pi)^{-p}g_s^{-1}\alpha^{\prime -\frac{(p+1)}{2}}$ is the brane tension ($g_s$ is the string coupling parameter, $\alpha^{\prime}$ is the inverse string tension). Also $F_{ab}$ is the field strength corresponding to the gauge field living in the worldvolume. If the supergravity background is generated by a stack of $N_c$ color branes, one must insert $N_f$ number of flavor branes such that $N_f \ll N_c$ (see Fig.\eqref{fig:intersecting branes}). This is called the probe limit, which ensures the backreaction of the flavor branes on the supergravity background can be neglected.
For $N_f$ number of flavor branes, the DBI action reads
\begin{align}
    S_{DBI}&=T_p N_f\int_{\mathcal{M}_{p+1}}d^{p+1}y \sqrt{-\det\Big[P(g_{ab}+B_{ab})+2\pi\alpha^{\prime}F_{ab}\Big]}\nonumber\\
    &=T_p N_f\int_{\mathcal{M}_{p+1}}d^{p+1}y \sqrt{-\det\Big[M_{ab}+F_{ab}\Big]}
\end{align}
where in the second line, we have defined $M_{ab}=P(g_{ab}+B_{ab})$ and have taken $2\pi\alpha^{\prime}=1$ for simplicity. Now, taking the delta variation of the above action, we obtain
\begin{equation}
    \delta S_{DBI}=T_p N_f \int d^{p+1}y\frac{1}{2}\sqrt{-\det (M_{ab})}M^{ab}\delta M_{ab}
\end{equation}
We can break $M^{ab}$ into a symmetric and anti-symmetric part. Thus, we can write $M^{ab}=\tilde S^{ab}+A^{ab}$, where $S^{ab}$ and $A^{ab}$ are the symmetric and anti-symmetric parts respectively. Taking this into account, we can recast the above equation in the following form
\begin{align}
    \delta S_{DBI}&=T_p N_f \int d^{p+1}y\frac{1}{2}\sqrt{-\det (M_{ab})}(\tilde S^{ab}+A^{ab})(\delta P[g_{ab}]+\delta F_{ab})\nonumber\\
    &=T_p N_f \int d^{p+1}y\frac{1}{2}\sqrt{-\det (M_{ab})}(\tilde S^{ab}\delta P[g_{ab}]+A^{ab}\delta F_{ab})~.
\end{align}
We would like to mention that, throughout our analysis, we implicitly assumed that the background has a vanishing NS-NS two-form. Therefore, we have set $B=0$. From the symmetry and anti-symmetry properties of the tensors, we find that the second term in the above equation vanishes, leaving only the first term, which reads
\begin{equation}
    \frac{\partial S_{DBI}}{\partial P[g_{ab}]}=T_p N_f \int d^{p+1}y\frac{1}{2}\sqrt{-\det (M_{ab})}\tilde S^{ab}
\end{equation}
We know that $M^{-1}M=I$, this gives
\begin{align}
    &(\tilde S^{ab}+A^{ab})(P[g_{ab}]+F_{ab})=I\nonumber\\
    &\implies \tilde S^{ab}+A^{ab}=[(P[g]+F)^{-1}]^{ab}\label{s+a}
\end{align}
Similarly
\begin{align}
    &(\tilde S^{ab}-A^{ab})(P[g_{ab}]-F_{ab})=I\nonumber\\
    &\implies \tilde S^{ab}-A^{ab}=[(P[g]-F)^{-1}]^{ab}\label{s-a}
\end{align}
Adding eq.\eqref{s+a} and eq.\eqref{s-a}, we get the following expressions for $\tilde S^{ab}$ and $A^{ab}$
\begin{align}
    &\tilde S^{ab}=\frac{1}{2}\Big[(P[g]+F)^{-1}+(P[g]-F)^{-1}\Big]^{ab}\label{tilde Sab}\\
    &A^{ab}=\frac{1}{2}\Big[(P[g]+F)^{-1}-(P[g]-F)^{-1}\Big]^{ab}\label{Aab}~.
\end{align}
We can rewrite the expressions for $(P[g_{ab}]+F_{ab})^{-1}$ and $(P[g_{ab}]-F_{ab})^{-1}$ in the following way
\begin{align}
    (P[g_{ab}]+F_{ab})^{-1}&=[(P[g]+F)^{-1}(P[g]-F)(P[g]-F)^{-1}]_{ab}\nonumber\\
    (P[g_{ab}]-F_{ab})^{-1}&=[(P[g]+F)^{-1}(P[g]+F)(P[g]-F)^{-1}]_{ab}~.
\end{align}
Hence, using the above identities, eq.\eqref{tilde Sab} can be simplified and written in the following form
\begin{align}
    \tilde S^{ab}&=\frac{1}{2}\Big[(P[g]+F)^{-1}\left\{(P[g]+F)+(P[g]-F)\right\}(P[g]-F)^{-1}\Big]^{ab}\nonumber\\
    &=\Big[(P[g]+F)^{-1}P[g](P[g]-F)^{-1}\Big]^{ab}
\end{align}
Now, taking the inverse of the above equation gives
\begin{equation}\label{Sab final}
    S_{ab}=P[g_{ab}]-\Big(F P[g]^{-1}F\Big)_{ab}
\end{equation}
where $S_{ab}$ is the inverse of $\tilde S^{ab}$. In the upcoming part, we will realise the importance of this tensor $S_{ab}$ while studying various scalar, vector, or spinor fluctuations on the flavor brane.\\
Till now, we were not bothered about the fluctuation of the flavor brane. The dynamical
fields of the DBI theory are the transverse scalars and the worldvolume gauge field. The fluctuations on the flavor brane result from the transverse and longitudinal oscillations of the open strings on the branes, respectively. Fluctuations around the classical saddle of scalar and gauge fields are given by
\begin{equation}
    \theta_i=\theta_i^{(0)}+\phi_i~,~~F_{ab}=F^{(0)}_{ab}+\mathcal{F}_{ab}
\end{equation}
where $\theta_i^{(0)}$ denotes the classical profile of the transverse scalars and $\phi_i$ represents the collective transverse fluctuations on the flavor brane. Similarly is the classical saddle point value of the field strength tensor and $\mathcal{F}_{ab}$ represent the collective fluctuations of the field strength on the probe flavor brane. The kinetic terms for these fluctuations take the following form
\begin{align}
    \mathcal{S}_{scalar}&=\frac{-\kappa}{2}\int d^{p+1}y \Big(\frac{\det g}{\det S}\Big)^{\frac{1}{4}}\sqrt{-\det S}\tilde S^{ab}\partial_a\phi^i\partial_b\phi^i\\
    \mathcal{S}_{vector}&=\frac{-\kappa}{4}\int d^{p+1}y \Big(\frac{\det g}{\det S}\Big)^{\frac{1}{4}}\sqrt{-\det S}\tilde S^{ab}\tilde S^{cd}\mathcal{F}_{ab}\mathcal{F}_{cd}
\end{align}
where $S$ is the determinant of the metric $S_{ab}$ given in eq.\eqref{Sab final}. It should be mentioned that the spinor fluctuations can be a supersymmetric counterpart of the DBI action,  which schematically consists of a standard Volkov-Akulov type term \cite{Akulov:1974xz}
\begin{equation}
    \mathcal{S}_{VA}=-N_f T_p\int d^{p+1}y e^{-\Phi}\sqrt{-\det(M+i\bar\psi\gamma\nabla \psi)}
\end{equation}
where the $\gamma$ metrices satisfy the following anticommutation relation with respect to $P[g]$
\begin{equation}
    \left\{\gamma_a,\gamma_b\right\}=2P[g_{ab}]~.
\end{equation}
Some key observations in this regard are in order. It is evident from the above expressions that the scalar and vector modes on the probe flavor branes perceive an effective metric, which is denoted by $S$. This metric is different from the background metric $g$ or the worldvolume-induced metric $P[g]$. This metric $S$ is the well-known open string metric (OSM) \cite{Seiberg:1999vs}. This OSM indeed describes the behavior of open string degrees of freedom moving within a background geometry that includes an antisymmetric 2-form field.
\subsection{Open string metric in \texorpdfstring{$(2+1)$}{(2+1)}-dimensions}
Now we will proceed to derive the OSM, taking AdS$_3$ spacetime as the background. We will also see the emergence of an event horizon in this OSM due to a specific choice of worldvolume gauge field.
The AdS$_3$ metric in the Poincaré coordinates is given by
\begin{equation}\label{AdS3 metric}
    ds^2=\frac{1}{z^2}\Big[-dt^2 +dx^2 +dz^2\Big]
\end{equation}
In order to introduce a horizon in the open string geometry, one needs to turn on an electric field. Therefore, one can choose the following gauge field along $x$ direction
\begin{equation}\label{gauge field}
    A_{x}=-Et + a_{x}(z)
\end{equation}
For space-filling branes, there is no pullback due to the insertion of the flavor branes. It states that $P[g_{ab}]=g_{ab}$. Hence, for the space-filling branes, the open string metric takes the following form
\begin{equation}
    S_{ab}=g_{ab}-\Big(F g^{-1}F\Big)_{ab}
\end{equation}
Now, for the background AdS$_3$ geometry in eq.\eqref{AdS3 metric} and the specific choice of the gauge field in eq.\eqref{gauge field}, we can find all the components of the open string metric. This reads
\begin{align}\label{S components}
    S_{tt}&=g_{tt}-F_{tx}g^{xx}F_{xt}=-\frac{1}{z^2}(1-E^2 z^4)\nonumber\\
    S_{xx}&=g_{xx}-F_{xt}g^{tt}F_{tx}-F_{xz}g^{zz}F_{zx}=\frac{1}{z^2}-E^2z^2+a^{\prime}_{x}(z)^2z^2\nonumber\\
    S_{zz}&=g_{zz}-F_{zx}g^{xx}F_{xz}=\frac{1}{z^2}+a^{\prime}_{x}(z)^2z^2\nonumber\\
    S_{tz}&=g_{tz}-F_{tx}g^{xx}F_{xz}=-E a^{\prime}_{x}(z)z^2=S_{zt}
\end{align}
Thus the final open string metric can be written as
\begin{align}
    S_{\mu\nu}dx^{\mu}dx^{\nu}&=S_{tt}dt^2+S_{xx}dx^2+S_{zz}dz^2+2S_{tz}dtdz\nonumber\\
    &=-\frac{1}{z^2}(1-E^2 z^4)dt^2+\Big(\frac{1}{z^2}-E^2z^2+a^{\prime}_{x}(z)^2z^2\Big)dx^2+\Big(\frac{1}{z^2}+a^{\prime}_{x}(z)^2z^2\Big)dz^2\nonumber\\
    &-2E a^{\prime}_{x}(z)z^2dzdt
\end{align}
The above OSM has cross terms. For the shake of simplicity, we will do the following coordinate transform
\begin{equation}
    \tau =t+f(z)
\end{equation}
where $f(z)$ is an arbitrary function of the inverse radial coordinate $z$. Under the coordinate transformation mentioned above we get
\begin{equation}\label{S diag}
    S_{\mu\nu}dx^{\mu}dx^{\nu}=S_{tt}(d\tau^2 -2f^{\prime}(z)d\tau dz +f^{\prime}(z)^2 dz^2)+S_{xx}dx^2+S_{zz}dz^2+2S_{tz}(d\tau dz -f^{\prime}(z)dz^2)~.
\end{equation}
In the above expression, we do not want any diagonal term. This gives us the following condition
\begin{equation}
    f^{\prime}(z)=\frac{S_{tz}}{S_{tt}}~.
\end{equation}
Putting the forms of $S_{tz}$ and $S_{tt}$ from eq.\eqref{S components}, we obtain
\begin{equation}\label{f prime}
    f^{\prime}(z)=\frac{-E z^4 a_{x}^{\prime}(z)}{(1-E^2 z^4)}~.
\end{equation}
Now we will derive an expression for $a_{x}^{\prime}(z)$. To do the same, we will have to compute the flavor current, which is given by
\begin{equation}
    j=\frac{\partial \mathcal{L}_{DBI}}{\partial a_{x}^{\prime}(z)}
\end{equation}
The Lagrangian of the DBI action is given by $\mathcal{L}_{DBI}=\sqrt{-\det(g+F)}$. Using the expressions for the background metric and gauge field from eq.(s)(\eqref{AdS3 metric},\eqref{gauge field}), one can compute the determinant inside the square root, which reads
\begin{equation}
    \det(g+F)=\det \begin{pmatrix}
-\frac{1}{z^2} & -E & 0 \\
E & \frac{1}{z^2} & a_{x}^{\prime}(z) \\
0 & -a_{x}^{\prime}(z) & \frac{1}{z^2}
\end{pmatrix}=-\frac{1}{z^6}\Big(1-E^2 z^4+z^4a_{x}^{\prime}(z)^2\Big)~.
\end{equation}
One can use the above equation for the determinant to find the DBI Lagrangian, which reads
\begin{equation}
    \mathcal{L}_{DBI}=\frac{1}{z^3}\sqrt{\Big(1-E^2 z^4+z^4a_{x}^{\prime}(z)^2\Big)}~.
\end{equation}
Therefore the flavor current is given by
\begin{equation}
    j=\frac{za_{x}^{\prime}(z)}{\sqrt{\Big(1-E^2 z^4+z^4 a_{x}^{\prime}(z)^2\Big)}}~.
\end{equation}
The above equation can be used to find the value of $a_{x}^{\prime}(z)$ in terms of the flavor current. Solving the above equation for $a_{x}^{\prime}(z)$ gives
\begin{equation}\label{aprime x}
    a_{x}^{\prime}(z)=\frac{j}{z}\sqrt{\frac{1-E^2 z^4}{1-j^2 z^2}}~.
\end{equation}
Now we will put this value for $a_{x}^{\prime}(z)$ in the expression of $f^{\prime}(z)$ (eq.\eqref{f prime}). This reads
\begin{equation}
    f^{\prime}(z)=\frac{-Ejz^3}{\sqrt{(1-E^2 z^4)(1-j^2 z^2)}}
\end{equation}
This gives the required coordinate transformation to diagonalise the OSM in $2+1$-dimensions. Hence, we finally get
\begin{equation}
    d\tau=dt-\frac{Ejz^3}{\sqrt{(1-E^2 z^4)(1-j^2 z^2)}}~.
\end{equation}
With this coordinate transformation relation in hand, we will now proceed further to calculate the components for the diagonal OSM. Therefore, components of the diagonal OSM are \footnote{A detailed calculation for obtaining these expressions for $\tilde S_{\tau\tau}$, $\tilde S_{xx}$ and $\tilde S_{zz}$ is given in appendix.}
\begin{align}
    \tilde S_{\tau\tau}&=-\frac{1}{z^2}(1-E^2z^4)\nonumber\\
    \tilde S_{xx}&=\frac{1}{z^2}+E\nonumber\\
    \tilde S_{zz}&=\frac{1}{z^2 (1-E z^2)}~.
\end{align}
Finally, the diagonal OSM in $2+1$-dimensions is given by
\begin{equation}\label{ds2 osm 2+1}
    ds^2_{osm}=\frac{1}{z^2}\Big[-(1-E^2 z^4)d\tau^2+(1+Ez^2)dx^2+\frac{dz^2}{(1-Ez^2)}\Big]~.
\end{equation}
In terms of the horizon radius $z_h$ the above metric can be written as follows
\begin{equation}
    ds^2_{osm}=-\frac{1}{z^2}\Big(1-\frac{z^4}{z_h^4}\Big)d\tau^2+\Big(\frac{1}{z^2}+\frac{1}{z_h^2}\Big)dx^2+\frac{dz^2}{z^2\Big(1-\frac{z^2}{z_h^2}\Big)}~.
\end{equation}
The OSM in presence of a background electric field in $3+1$ and $4+1$-dimensions are also given in \cite{Banerjee:2020gyv,Banerjee:2016qeu}, which are given below
\begin{align}
ds^2_{(4)}
&=
\frac{1}{z^2}\left[
-(1-E^2 z^4)\,dt^2
+\frac{dz^2}{1-E^2 z^4}
+\left(dx_1^2+dx_2^2\right)
\right], \\[1em]
ds^2_{(5)}
&=
\frac{1}{z^2}\left[
-(1-E^2 z^4)\,dt^2
+\frac{dz^2}{1-E^3 z^6}
+\frac{1-E^2 z^4}{1-E^3 z^6}\,dx_1^2
+\left(dx_2^2+dx_3^2\right)
\right].
\end{align}
In the upcoming sections, we will use these OSMs to compute HEE, HMI, EWS, EN, etc. We would like to mention that we have done all the calculations with OSMs in $(2+1)$ and $(3+1)$-dimensions only.
\section{Holographic entanglement entropy}\label{Sec:3}
In this section, we will briefly revisit the calculation of the HEE for the OSMs in $(2+1)$ and $(3+1)$-dimensions. These computations were first done in \cite{Banerjee:2020gyv} for OSMs in $(2+1)$, $(3+1)$ and $(4+1)$-dimensions. Here, we provide a more detailed calculation for the HEE. The results of this section will be useful later in order to find the mutual information and entanglement negativity of the boundary theory in a holographic manner. Before proceeding further, we must discuss a little bit about the holographic calculation of entanglement entropy. The holographic entanglement entropy (HEE) is the gravity dual of the entanglement entropy in a conformal field theory (CFT) situated at the boundary. The famous Ryu-Takayanagi (RT) formula \cite{Ryu:2006bv,Ryu:2006ef} is used to compute the HEE for a static boundary CFT. The RT formula connects the entanglement entropy of a strongly coupled conformal field theory (CFT) to the area of a static minimal surface in the bulk spacetime with codimension two. This minimal surface is commonly referred to as the RT surface. According to the RT formula, the HEE of a subsystem $A$ leaving on a $d$-dimensional boundary is given by
\begin{equation}
    S_{HEE}(A)=\frac{1}{4G_N^{(d+1)}}[Area(\Gamma^A_{min})]
\end{equation}
where $G_N^{(d+1)}$ is the Newton’s gravitational constant in $(d+1)$-dimensions and $\Gamma^A_{min}$ is the $(d-1)$-dimensional static minimal surface such that $\partial\Gamma^A_{min}=\partial A$.
\subsection{HEE for (2+1)-dimensional open string metric}
\begin{figure}
    \centering
    \includegraphics[width=0.6\linewidth]{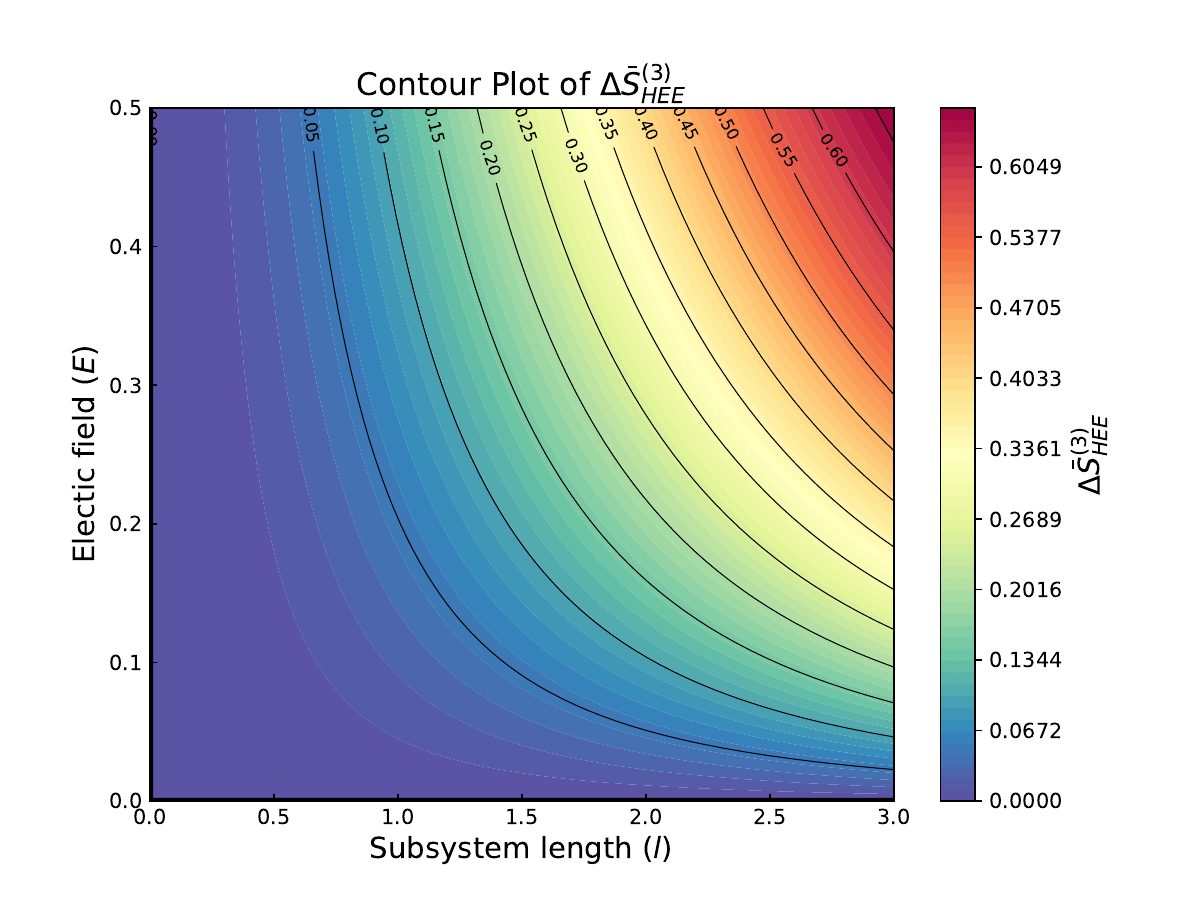}
    \caption{A contour plot depicting the variation of the change in HEE, denoted as $\Delta \bar{S}^{(4)}_{HEE}$, in a $(2+1)$-dimensional OSM background, as a function of two independent parameters: the applied electric field $E$ and the width $l$ of the boundary subsystem.}
    \label{fig:S3contour}
\end{figure}

We will now compute the HEE for the OSM in $(2+1)$-dimensions. The OSM in $(2+1)$-dimensions is given by \cite{Banerjee:2016qeu,Banerjee:2020gyv}
\begin{equation}
    ds^2_{(3)}=\frac{1}{z^2}\Big[-(1-E^2 z^4)d\tau^2+(1+Ez^2)dx^2+\frac{dz^2}{(1-Ez^2)}\Big]~.
\end{equation}
For a static RT surface, $d\tau =0$, and we will also choose the parameterisation $z=z(x)$. These choices give us the following induced metric
\begin{equation}
    dS^2_{(3)ind}=\frac{1}{z^2}\Bigg[(1+Ez^2)dx^2+\frac{\Big(\frac{dz}{dx}\Big)^2}{(1-Ez^2)}dx^2\Bigg]~.
\end{equation}
Using the above induced metric, one can easily compute the RT area functional, which reads
\begin{equation}\label{A in dx}
    \mathcal{A}_{(3)}=\int_{-l/2}^{l/2}dx\frac{1}{z}\sqrt{(1+Ez^2)+\frac{z^{\prime 2}}{(1-Ez^2)}}
\end{equation}
where $z^{\prime}=\frac{dz}{dx}$. From the above equation, we can easily identify the Lagrangian to be 
\begin{equation}
    \mathcal{L}_{(3)}=\frac{1}{z}\sqrt{(1+Ez^2)+\frac{z^{\prime 2}}{(1-Ez^2)}}
\end{equation}
Hence, the canonical momentum along the $z$ direction is given by
\begin{equation}
    \mathcal{P}_{z}=\frac{\partial\mathcal{L}_{(2)}}{\partial z^\prime}=\frac{(\frac{z^\prime}{z})}{(1-Ez^2)\sqrt{(1+Ez^2)+\frac{z^{\prime 2}}{(1-Ez^2)}}}
\end{equation}
Therefore, the Hamiltonian is given by the following expression
\begin{equation}
    \mathcal{H}=\mathcal{P}_z z^{\prime}-\mathcal{L}_{(2)}=\frac{-(1+Ez^2)}{z\sqrt{(1+Ez^2)+\frac{z^{\prime 2}}{(1-Ez^2)}}}~.
\end{equation}
From the property of the RT surface, we know that at the turning point ($z=z_t$), $z^{\prime}(x)=0$. Hence, at the turning point, the Hamiltonian is given by
\begin{equation}\label{H at zt}
    \mathcal{H}\mid_{z=z_t}=-\frac{\sqrt{1+E z_{t}^2}}{z_t}=b
\end{equation}
where $b$ is a constant. From the conservation of $\mathcal{H}$, we can write
\begin{equation}\label{b val}
    \frac{-(1+Ez^2)}{z\sqrt{(1+Ez^2)+\frac{z^{\prime 2}}{(1-Ez^2)}}}=b~.
\end{equation}
Rearranging the above equation leads to the following
\begin{equation}
    x^{\prime 2}=\frac{b^2 z^2}{(1-Ez^2)\Big[(1+Ez^2)^2-b^2z^2 (1+Ez^2)\Big]}
\end{equation}
where $x^{\prime}=\frac{dx}{dz}$.\\
As we already know the value of the constant $b$ from eq,\eqref{H at zt}. Substituting that value of $b$ into the above equation gives
\begin{equation}\label{x prime}
    x^{\prime 2}=\frac{\Big(\frac{z}{z_t}\Big)^2}{(1-E^2 z^4)\Bigg[\Big(\frac{1+Ez^2}{1+Ez_t^2}\Big)-\frac{z^2}{z_t^2}\Bigg]}~.
\end{equation}
The above expression will be useful in order to find an expression of the turning point ($z_t$) in terms of the boundary subsystem size ($l$). Integrating both sides, we get
\begin{equation}
    \int_{0}^{\frac{l}{2}}dx=\frac{l}{2}=\int_{0}^{z_t}\frac{\Big(\frac{z}{z_t}\Big)}{\sqrt{1-E^2 z^4}\sqrt{\Big(\frac{1+Ez^2}{1+Ez_t^2}\Big)-\frac{z^2}{z_t^2}}}dz
\end{equation}
In order to evaluate the integral, let us do a variable transformation $y=\frac{z}{z_t}$. Under this change of variables, the above integral can be recast into the following form
\begin{equation}\label{l in y coord}
    \frac{l}{2}=z_t\int_{0}^{1}\frac{y\sqrt{1+Ez_t^2}}{\sqrt{1-E^2 z_t^4 y^4}\sqrt{1-y^2}}dy
\end{equation}
Similarly, we can use eq.\eqref{x prime} to transform the integral for the area functional (eq.\eqref{A in dx}), this reads
\begin{equation}
    \mathcal{A}_{(3)}=2\int_{\epsilon}^{z_t}\frac{\sqrt{1+Ez^2}}{z\sqrt{(1-Ez^2)(1+Ez^2-b^2 z^2)}}dz
\end{equation}
where $\epsilon$ is the UV cutoff for the RT surface and is situated close to the boundary. \\
Now we will substitute the value of the constant parameter $b$ from eq.\eqref{b val} in the above expression for the area functional, which reads
\begin{equation}
    \mathcal{A}_{(3)}=2\int_{\epsilon}^{z_t}\frac{dz}{z}\frac{\sqrt{1+Ez^2}}{\sqrt{1-\frac{z^2}{z_t^2}}\sqrt{1-Ez^2}}~.
\end{equation}
Again transforming the above integral in $y$ coordinate gives the following
\begin{equation}\label{A2 in y coord}
    \mathcal{A}_{(3)}=2\int_{\epsilon/z_t}^{1}dy \frac{\sqrt{1+Ez_t^2 y^2}}{y\sqrt{1-y^2}\sqrt{1-Ez_t^2 y^2}}~.
\end{equation}
Now we will compute the turning point in terms of the boundary subsystem length. To do the same, we will use eq.\eqref{l in y coord}, which reads
\begin{equation}
    \frac{l}{2}=z_t+\frac{E}{2}z_t^3+\frac{17E^2}{120}z_t^5+\mathcal{O}(E^3)+\dots
\end{equation}
While obtaining the above equation, we have expanded the integrand in eq.\eqref{l in y coord} by considering the electric field $E$ as a small perturbative parameter. For the rest of the paper, we have neglected terms which are higher than $\mathcal{O}(E^2)$. A series inversion of the above expression gives an approximate value of the turning point ($z_t$) in terms of the subsystem length ($l$), which reads
\begin{equation}\label{zt in l}
    z_t\approx\frac{l}{2}\Bigg[1-\frac{E}{8}l^2+\frac{73E^2}{1920}l^4\Bigg]~.
\end{equation}
Using the above expression, we can evaluate the integral for the area functional. By performing a series expansion of the integrand in eq.\eqref{A2 in y coord} for small values of the electric field, we get
\begin{align}\label{A2 in I}
    \mathcal{A}_{(3)}&\approx 2\int_{\epsilon/z_t}^{1}\frac{dy}{y\sqrt{1-y^2}}\Big[1+Ez_t^2y^2+\frac{1}{2}E^2 z_t^4 y^4\Big]\nonumber\\
    &=2\Big[I_0+Ez_t^2 I_2+\frac{1}{2}E^2 z_t^4 I_4\Big]
\end{align}
where we have defined
\begin{align}
    I_0&=\int_{\epsilon/z_t}^{1}\frac{dy}{y\sqrt{1-y^2}}=-\frac{1}{2}\log\Bigg[\frac{1-\sqrt{1-\frac{\epsilon^2}{z_t^2}}}{1+\sqrt{1-\frac{\epsilon^2}{z_t^2}}}\Bigg]\approx\log\Big(\frac{2z_t}{\epsilon}\Big)\nonumber\\
    I_2&=\int_{\epsilon/z_t}^{1}\frac{ydy}{\sqrt{1-y^2}}=\sqrt{1-\frac{\epsilon^2}{z_t^2}}\approx 1\nonumber\\
    I_4&=\int_{\epsilon/z_t}^{1}\frac{y^3dy}{\sqrt{1-y^2}}=\frac{1}{3}\Big(2+\frac{\epsilon^2}{z_t^2}\Big)\sqrt{1-\frac{\epsilon^2}{z_t^2}}\approx \frac{2}{3}~.
\end{align}
After substituting these values of $I_0$, $I_2$ and $I_4$ in eq.\eqref{A2 in I}, we obtain
\begin{equation}\label{A in zt}
    \mathcal{A}_{(3)}=2\Bigg[\log\Big(\frac{2z_t}{\epsilon}\Big)+Ez_t^2+\frac{1}{3}E^2 z_t^4\Bigg]~.
\end{equation}
Finally, we can use the above equation to determine the area functional ($\mathcal{A}_{(2)}$) in terms of the subsystem length ($l$). In order to do this, we will put the expression for $z_t$ (eq.\eqref{zt in l}) in eq.\eqref{A in zt}, which approximately gives
\begin{equation}
    \mathcal{A}_{(3)}\approx 2\Bigg[\log\Big(\frac{l}{\epsilon}\Big)+\frac{1}{8}El^2-\frac{11}{960}E^2 l^4\Bigg]
\end{equation}
where we have neglected terms of $\mathcal{O}(E^3)$ and consecutive higher orders in $E$. Now, by applying the RT formula, one can easily calculate the HEE for the boundary subsystem of length $l$, which reads
\begin{equation}\label{HEE 2+1}
    S^{(3)}_{HEE}(l)\approx \frac{1}{2 G_3}\Bigg[\log\Big(\frac{l}{\epsilon}\Big)+\frac{1}{8}El^2-\frac{11}{960}E^2 l^4\Bigg]
\end{equation}
Hence, the change in HEE from pure AdS$_3$ geometry due to the application of an electric field in the flavor brane world volume is given by
\begin{equation}
    \Delta S^{(3)}_{HEE}(l)=\frac{El^2}{4G_3}\Bigg[\frac{1}{4}-\frac{11}{480}El^2\Bigg]~.
\end{equation}
For simplicity, one can also define 
\begin{equation}
    \Delta \bar S^{(3)}_{HEE}(l)=4G_3\Delta  S^{(3)}_{HEE}(l)=El^2\Bigg[\frac{1}{4}-\frac{11}{480}El^2\Bigg]~.
\end{equation}
We have also done a contour plot for $\Delta \bar S^{(3)}_{HEE}(l)$ with respect to two independent parameters (subsystem length $l$ and applied electric field $E$) in Fig.\eqref{fig:S3contour}.
\subsection{HEE for (3+1)-dimensional open string metric}
\begin{figure}[t]
    \centering
    \includegraphics[width=0.6\linewidth]{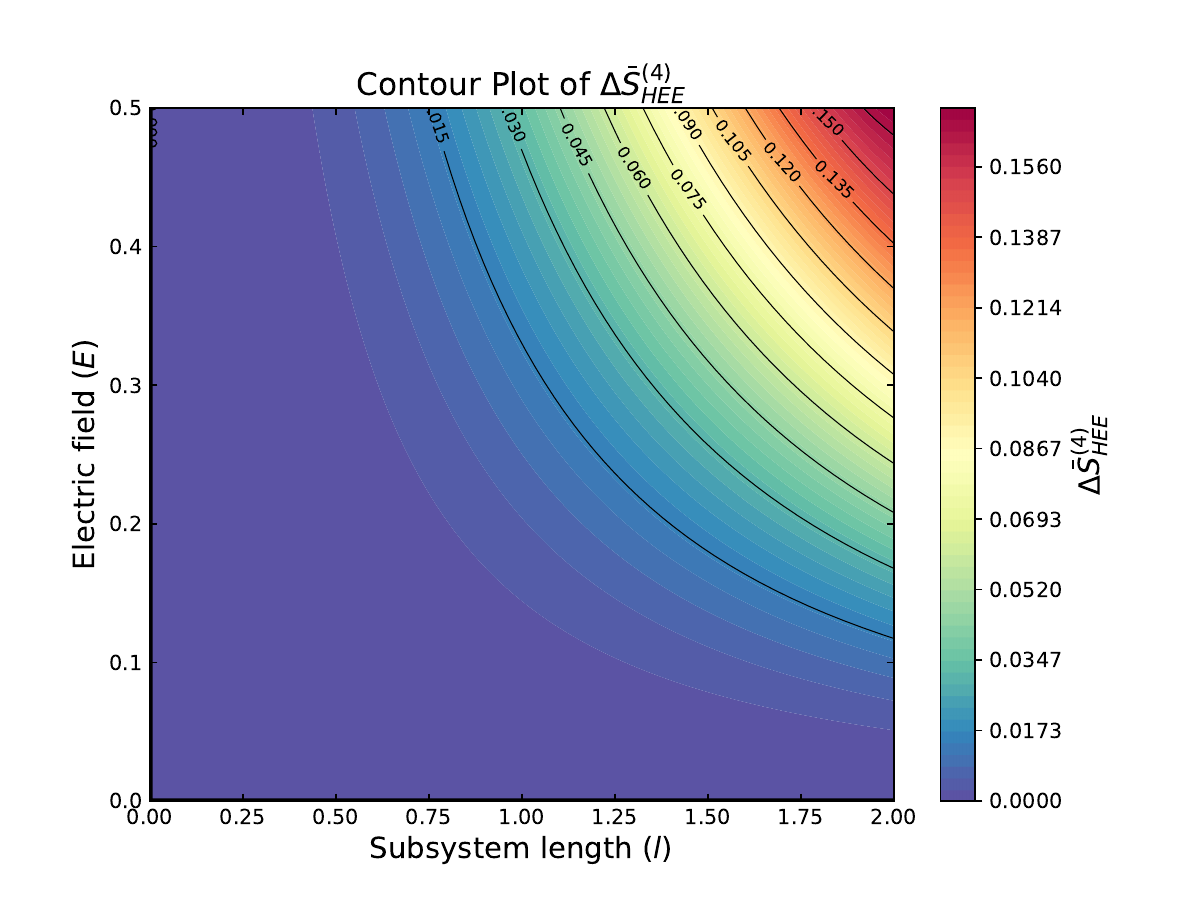}
    \caption{Contour plot illustrating the variation of the change in HEE of $(3+1)$-dimensional OSM $\Delta \bar S^{(4)}_{HEE}$ as a function of two independent parameters: applied electric field $E$ and boundary subsystem width $l$.}
    \label{fig:S4contour}
\end{figure}

In this subsection, we will focus on a brief re-derivation of the HEE for ($3+1$)-dimensional OSM as mentioned in \cite{Banerjee:2016qeu,Banerjee:2020gyv} 
The $(3+1)$-dimensional OSM is given by
\begin{equation}
    ds^2_{(4)}=
\frac{1}{z^2}\left[
-(1-E^2 z^4)\,dt^2
+\frac{dz^2}{1-E^2 z^4}
+\left(dx_1^2+dx_2^2\right)
\right]
\end{equation}
Now let us choose a strip-like boundary subsystem with $-l/2\leq x_1\leq l/2$ and $0\leq x_2 \leq L$. 
In order to find out the static RT surface for this kind of boundary subsystem, we first need to find the induced metric with $d\tau =0$ and $z=z(x_1)$, which reads
\begin{equation}
    ds^2_{(4)ind}=\frac{1}{z^2}\Bigg[\Big(1+\frac{z^{\prime 2}(x_1)}{1-E^2 z^4}\Big)dx_1^2 +dx_2^2\Bigg]~.
\end{equation}
The above equation gives the following RT area functional
\begin{equation}\label{A4}
    \mathcal{A}^{(4)}=\int_{0}^{L}dx_2\int_{-l/2}^{l/2}\frac{dx_1}{z^2}\sqrt{1+\frac{z^{\prime 2}(x_1)}{1-E^2 z^4}}=L\int_{-l/2}^{l/2}\frac{dx_1}{z^2}\sqrt{1+\frac{z^{\prime 2}(x_1)}{1-E^2 z^4}}
\end{equation}
We can identify the Lagrangian from the above expression of the area functional, which reads
\begin{equation}
    \mathcal{L}_{(4)}=\frac{1}{z^2}\sqrt{1+\frac{z^{\prime 2}(x_1)}{1-E^2 z^4}}~.
\end{equation}
The canonical momentum in the $z$ direction is given by
\begin{equation}
    \mathcal{P}^{(4)}_z=\frac{\partial \mathcal{L}_{(4)}}{\partial z^{\prime}}=\frac{z^{\prime}}{z^2(1-E^2 z^4)\sqrt{1+\frac{z^{\prime 2}}{1-E^2 z^4}}}
\end{equation}
Using this expression of $\mathcal{P}^{(4)}_z$, one can easily calculate the following Hamiltonian
\begin{equation}
    \mathcal{H}^{(4)}=\mathcal{P}^{(4)}_z z^{\prime}-\mathcal{L}_{(4)}=-\frac{1}{z^2 \sqrt{1+\frac{z^{\prime 2}}{1-E^2 z^4}}}
\end{equation}
At the turning point $z=z_*$, the above expression becomes
\begin{equation}
    \mathcal{H}^{(4)}\mid_{z=z_*}=-\frac{1}{z_*^2}~.
\end{equation}
The conservation of $\mathcal{H}^{(4)}$ suggests $\mathcal{H}^{(4)}=\mathcal{H}^{(4)}\mid_{z=z_*}$, this gives
\begin{equation}\label{x prime 4}
    x^{\prime 2}=\frac{1}{\sqrt{(1-E^2 z^4)}\sqrt{(\frac{z_*}{z})^4-1}}~.
\end{equation}
The above equation can be rearranged and used to rewrite the subsystem length ($l$) in terms of the turning point ($z_*$). This gives
\begin{equation}
    \frac{l}{2}=z_{*}\int_{0}^{1}dy \frac{y^2}{\sqrt{1-y^4}\sqrt{1-E^2 z_{*}^4 y^4}}
\end{equation}
where $y=\frac{z}{z_*}$.\\
The above integral can be evaluated perturbatively in a regime $Ez_*^2<1$. In this regime one gets
\begin{equation}
    \frac{l}{2}\approx z_*b_0\Big(1+\frac{3}{10}E^2 z_*^4+\frac{7}{40}E^4 z_*^8\Big)
\end{equation}
where $b_0=\frac{\sqrt{\pi}\Gamma(3/4)}{\Gamma(1/4)}$.\\
Inverting the above series and keeping terms up to $\mathcal{O}(E^4)$, we obtain
\begin{equation}\label{zt 3+1}
     z_* =\tilde z_* \Big(1-\frac{3}{10}E^2 \tilde z_*^4 +\frac{11}{40}E^4 \tilde z_*^8\Big)~.
\end{equation}
where $\tilde z_* =\frac{l}{2b_0}$.\\
The above relation of the turning point in terms of the subsystem length will be useful to determine the HEE in terms of the boundary subsystem length $l$. Now we will use eq.\eqref{x prime 4} and rewrite the integral for the area functional in eq.\eqref{A4}, which reads
\begin{align}
    \mathcal{A}^{(4)}&=2L\int_{\epsilon}^{z_*}\frac{dz}{z^2}\frac{1}{\sqrt{1-E^2 z^4}}\frac{1}{\sqrt{1-(\frac{z}{z_*})^4}}\nonumber\\&=2L\int_{\epsilon/z_*}^{1}\frac{dy}{z_* y^2}\frac{1}{\sqrt{1-E^2 z_*^4 y^4}}\frac{1}{\sqrt{1-y^4}}~.
\end{align}
In the second line, we introduced the variable substitution $y = \frac{z}{z_*}$. Now we will move forward to evaluate the integral in the regime $Ez_*^2 <1$. Hence, we get the following perturbative expansion for $\mathcal{A}^{(4)}$
\begin{equation}\label{A4 int series}
    \mathcal{A}^{(4)}=\frac{2L}{z_*}\Big[\tilde I_0+\frac{E^2 z_*^4}{2} \tilde I_1+\frac{3E^4 z_*^8}{8}\tilde I_2\Big]
\end{equation}
where
\begin{align}
    \tilde I_0&=\int_{\epsilon/z_t}^{1}\frac{dy}{y^2\sqrt{1-y^4}}=-\frac{{}_2F_1\!\left(-\frac{1}{4},\frac{1}{2};\frac{3}{4};y^4\right)}{y}\Bigm|_{\frac{\epsilon}{z_*}}^{1}\approx\Big(\frac{z_*}{\epsilon}-b_0\Big)\nonumber\\
    \tilde I_1&=\int_{\epsilon/z_t}^{1}\frac{y^2 dy}{\sqrt{1-y^4}}=\frac{y^3}{3}{}_2F_1(\frac{1}{2},\frac{3}{4};\frac{7}{4};y^4)\Bigm|_{\frac{\epsilon}{z_*}}^{1}\approx b_0\nonumber\\
    \tilde I_2&=\int_{\epsilon/z_t}^{1}\frac{y^6 dy}{\sqrt{1-y^4}}=\frac{y^3}{5}
\left(
-\sqrt{1-y^4}
+ {}_2F_1\!\left(
\frac{1}{2},\frac{3}{4};\frac{7}{4};y^4
\right)
\right)\Bigm|_{\frac{\epsilon}{z_*}}^{1}\approx \frac{3}{5}b_0~.
\end{align}
Now using these expressions for $\tilde I_0$, $\tilde I_1$ and $\tilde I_2$ in eq.\eqref{A4 int series}, we finally get
\begin{align}
    \mathcal{A}^{(4)}&=2\Big(\frac{L}{\epsilon}-\frac{L b_0}{\tilde z_*}\Big)+2L \Big(\frac{L E^2 b_0}{5}\tilde z_*^3-\frac{LE^4 b_0}{25}\tilde z_*^7\Big)\nonumber\\
    &=2\Big(\frac{L}{\epsilon}-\frac{2L b_0^2}{l}\Big)+2L \Big(\frac{E^2 }{40 b_0^2} l^3-\frac{E^4 }{3200 b_0^6}l^7\Big)~.
\end{align}
With the area functional in hand, one can now use the RT formula to evaluate the HEE for the $(3+1)$-dimensional OSM, which reads
\begin{equation}\label{HEE 3+1 final}
    S^{(4)}_{HEE}=\frac{L}{4G_4}\Bigg[\Big(\frac{2}{\epsilon}-\frac{4 b_0^2}{l}\Big)+ \Big(\frac{E^2 }{20 b_0^2} l^3-\frac{E^4 }{1600 b_0^6}l^7\Big)\Bigg]
\end{equation}
Clearly the first term in the above expression is the HEE of pure AdS$_4$ geometry. Hence, the change in HEE from pure AdS$_4$ spacetime is given by
\begin{equation}\label{change HEE 3+1}
    \Delta S^{(4)}_{HEE}=\frac{L}{4G_4}\Big[ \frac{E^2 }{20 b_0^2} l^3-\frac{E^4 }{1600 b_0^6}l^7\Big]
\end{equation}
Here we can also define
\begin{equation}\label{change HEE 3+1 bar}
    \Delta \bar S^{(4)}_{HEE}(l)=\frac{4G_4}{L}\Delta S^{(4)}_{HEE}=\Big[ \frac{E^2 }{20 b_0^2} l^3-\frac{E^4 }{1600 b_0^6}l^7\Big]
\end{equation}
We have also prepared a contour plot for the expression of $\Delta \bar S^{(4)}_{HEE}(l)$ in eq.\eqref{change HEE 3+1 bar} as a function of two independent parameters that is subsystem length ($l$) and applied electric field ($E$), in Fig.\eqref{fig:S4contour}. 
\section{Holographic mutual information and EWCS}\label{Sec:4}
In this section, we will study the mutual information and entanglement wedge cross section for a parallel strip-like boundary subsystem with ($2+1$) and ($3+1$)-dimensional bulk open string geometries with a horizon induced by world-volume electric fields of flavor branes.  
\subsection{HMI and EWCS in $(2+1)$-dimensions}\label{subsec: 4.1}
\begin{figure}
    \centering
    \includegraphics[width=0.65\linewidth]{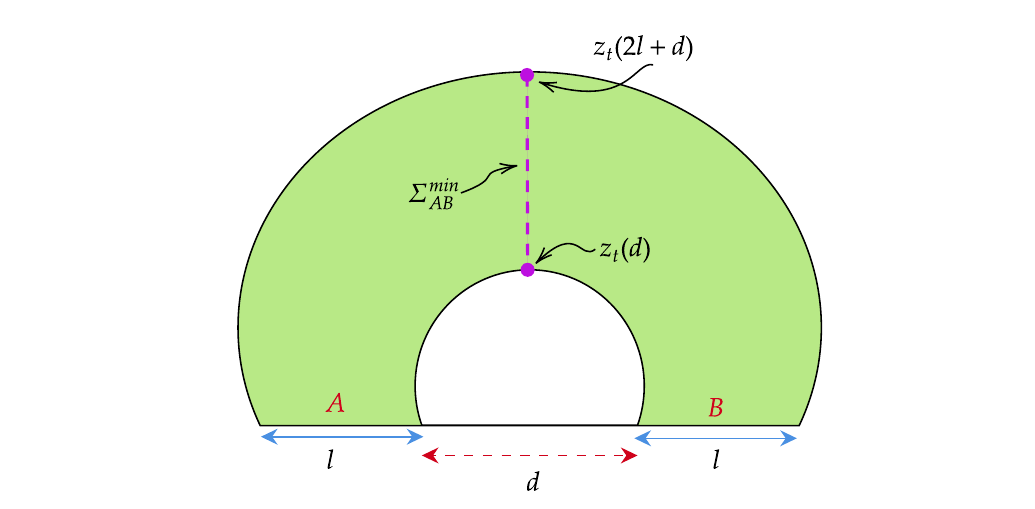}
    \caption{A pictorial representation of two thin strip-like boundary subsystems ($A$ and $B$) of equal width $l$ and separated by a distance $d$. The shaded region in green denotes the entanglement wedge $M_{AB}$, and the purple dashed line indicates the entanglement wedge cross-section $\Sigma^{min}_{AB}$. The two points $z_t(2l+d)$ and $z_t(d)$ indicates the turning point corresponding to the RT surfaces $\Gamma^{min}_{l\cup l}$ and $\Gamma^{min}_{d}$ respectively.}
    \label{fig:ewcs diagram}
\end{figure}

The HEE for a $(2+1)$-dimensional OSM is given by eq.\eqref{HEE 2+1}. This relation can be used to compute the mutual information between two parallel strip-like boundary subsystems of equal width ($l$) in a holographic manner. The HMI between two parallel boundary subsystems ($A$ and $B$) of width $l$ and separated by a distance $d$ is given by
\begin{align}\label{HMI main formula}
    I^{(2)}(A:B)&=S^{(2)}_{HEE}(A)+S^{(2)}_{HEE}(B)-S^{(2)}_{HEE}(A\cup B)\nonumber\\
    &=2S^{(2)}_{HEE}(l)-S^{(2)}_{HEE}(2l+d)-S^{(2)}_{HEE}(d)~.
\end{align}
using the above expression along with eq.\eqref{HEE 2+1}, we obtain
\begin{align}
    I^{(2)}(A:B)&=\frac{1}{4G_{3}}\Big[4\log (l)-2\log (2l+d)-2\log (d)\Big]+\frac{E}{16G_3}\Big[2l^2-(2l+d)^2-d^2\Big]\nonumber\\
    &-\frac{11E^2}{1920 G_3}\Big[2l^4-(2l+d)^4-d^4\Big]
\end{align}
From the above equation, it is clear that the mutual information between two subsystems is independent of the UV cutoff $\epsilon$.\\
We will now proceed to discuss the construction of the holographic counterpart of the entanglement of purification (EoP), which is known as the entanglement wedge cross-section (EWCS). In \cite{Takayanagi:2017knl,Nguyen:2017yqw,Jokela:2019ebz,BabaeiVelni:2019pkw} the authors have proposed a systematic way to compute EWCS following the "$E_P=E_W$". duality. We will start with two parallel boundary subsystems ($A$ and $B$) of equal width $l$ and separated by a distance $d$. Let us also consider $\Gamma^{min}_{A}$, $\Gamma^{min}_{B}$ and $\Gamma^{min}_{A\cup B}$ are the RT surfaces corresponding to the subsystems $A$, $B$ and $A\cup B$ respectively. For two non-overlapping subsystems ($A$ and $B$), we know that $A\cap B =0$. Now the bulk region surrounded by $A\cup B$ and $\Gamma^{min}_{A\cup B}$ is called the entanglement wedge ($M_{AB}$). The following boundary characterises the co-dimension-$0$ domain of the entanglement wedge
\begin{equation}
    \partial M_{AB}=A\cup B \cup \Gamma^{min}_{A\cup B}=\bar\Gamma_{A}\cup \bar \Gamma_{B}
\end{equation}
where $\bar\Gamma_{A}=A\cup \Gamma^{A}_{A\cup B}$ and $\bar\Gamma_{B}=B\cup \Gamma^{B}_{A\cup B}$. In order to obtain the above equation, we have also used $\Gamma^{min}_{A\cup B}=\Gamma^{A}_{A\cup B}\cup \Gamma^{B}_{A\cup B}$. As the boundary of the entanglement wedge is divided into two parts ($\bar\Gamma_{A}$ and $\bar\Gamma_{B}$), one can define holographic entanglement entropies $S(\rho_{\bar\Gamma_{A}})$ and $S(\rho_{\bar\Gamma_{B}})$ for $\bar\Gamma_{A}$ and $\bar\Gamma_{B}$ respectively. In order to compute the entanglement entropy between $\bar\Gamma_{A}$ and $\bar\Gamma_{B}$, we need to find the minimal surface ($\Sigma^{min}_{AB}$) which devides the entanglement wedge $M_{AB}$ into two parts (for clear visual understanding see Fig.\eqref{fig:ewcs diagram}). The minimal surface $\Sigma^{min}_{AB}$ is found such that it obeys the following relations
\begin{align}
    &(i)~~\partial\Sigma^{min}_{AB}=\partial \bar\Gamma_{A}=\partial\bar\Gamma_{B}\nonumber\\&(ii)~~\Sigma^{min}_{AB}~\text{is homologous to }\bar\Gamma_{A}~\text{inside}~M_{AB}.
\end{align}
There are infinitely many possible ways to divide the entanglement wedge ($M_{AB}$) into two parts. Among all possible choices, the one with the minimum area is useful to compute the entanglement between $\bar\Gamma_{A}$ and $\bar\Gamma_{B}$. This defines a quantity which
we call the entanglement wedge cross-section, which is given by
\begin{equation}
    E_{W}(\rho_{AB})=\min_{\bar\Gamma_{A}\subset \Gamma^{min}_{AB}}\Bigg[\frac{Area(\Sigma^{min}_{AB})}{4G_{d+1}}\Bigg]~.
\end{equation}
\noindent Now we will proceed to compute the EWCS between the two parallel strip-like boundary subsystems of width $l$ using the above formula. To compute the EWCS, we need to find the induced metric on a static patch at a constant boundary coordinate $x$. Therefore, setting $dt=dx=0$ in eq.\eqref{ds2 osm 2+1}, we obtain
\begin{equation}
    ds^2_{ind}=\frac{1}{z^2}\frac{dz^2}{1-Ez^2}
\end{equation}
Thus, the EWCS is given by
\begin{equation}
    E^{(2)}_{W}=\frac{1}{4 G_3}\int_{z_t(d)}^{z_t(2l+d)}dz\frac{1}{z \sqrt{1-Ez^2}}~.
\end{equation}
We shall expand the above integrand for small values of the applied electric field $E$, which gives
\begin{align}\label{EWCS in zt}
    E^{(3)}_{W}&\approx\frac{1}{4 G_3}\int_{z_t(d)}^{z_t(2l+d)}dz\Big[\frac{1}{z}+\frac{E}{2}z+\frac{3E^2}{8}z^3\Big]\nonumber\\
    &=\frac{1}{4 G_3}\Big[\ln z_t(2l+d)-\ln z_t(d)\Big]+\frac{1}{4G_3}\frac{E}{4}\Big[z_t(2l+d)^2-z_t(d)^2\Big]+\frac{3E^2}{128G_3}\Big[z_t(2l+d)^4-z_t(d)^4\Big]
\end{align}
In the above equation, we have neglected the terms higher than $\mathcal{O}(E^2)$. Now we will proceed further to evaluate the EWCS in terms of the boundary subsystem width ($l$). In order to do this, we will use the functional form of the turning point ($z_t$) from eq.\eqref{zt in l} in eq.\eqref{EWCS in zt}. This gives 
\begin{align}
    E^{(3)}_{W}&=\frac{1}{4G_3}\Bigg[\left\{\ln (2l+d)-\ln (d)\right\}-\frac{E}{2}(dl+l^2)+\frac{29E^2}{120}(d^3l+3d^2l^2+4dl^3+2l^4)\Bigg]\nonumber\\
    &+\frac{E}{16G_3}\Bigg[\frac{1}{4}\left\{(2l+d)^2-d^2\right\}+\frac{E}{4}\left\{\frac{d^4}{4}-\frac{(2l+d)^4}{4}\right\}\Bigg]+\frac{3E^2}{128G_3}\Bigg[\frac{1}{16}\left\{(2l+d)^4-d^4\right\}\Bigg]~.
\end{align}
After doing a little bit of algebra, the above equation can be further simplified and recast in the following form
\begin{equation}
    \bar E^{(3)}_{W}=\left\{\ln (2l+d)-\ln (d)\right\}-\frac{E}{4}l(d+l)+\frac{157E^2}{960}l(d+l)(d^2+2dl+2l^2)
\end{equation}
where $\bar E^{(3)}_{W}=4G_3 E^{(3)}_{W}$.\\
\begin{figure}
     \centering
     \begin{subfigure}[t]{0.45\textwidth}
         \centering
         \includegraphics[width=\textwidth]{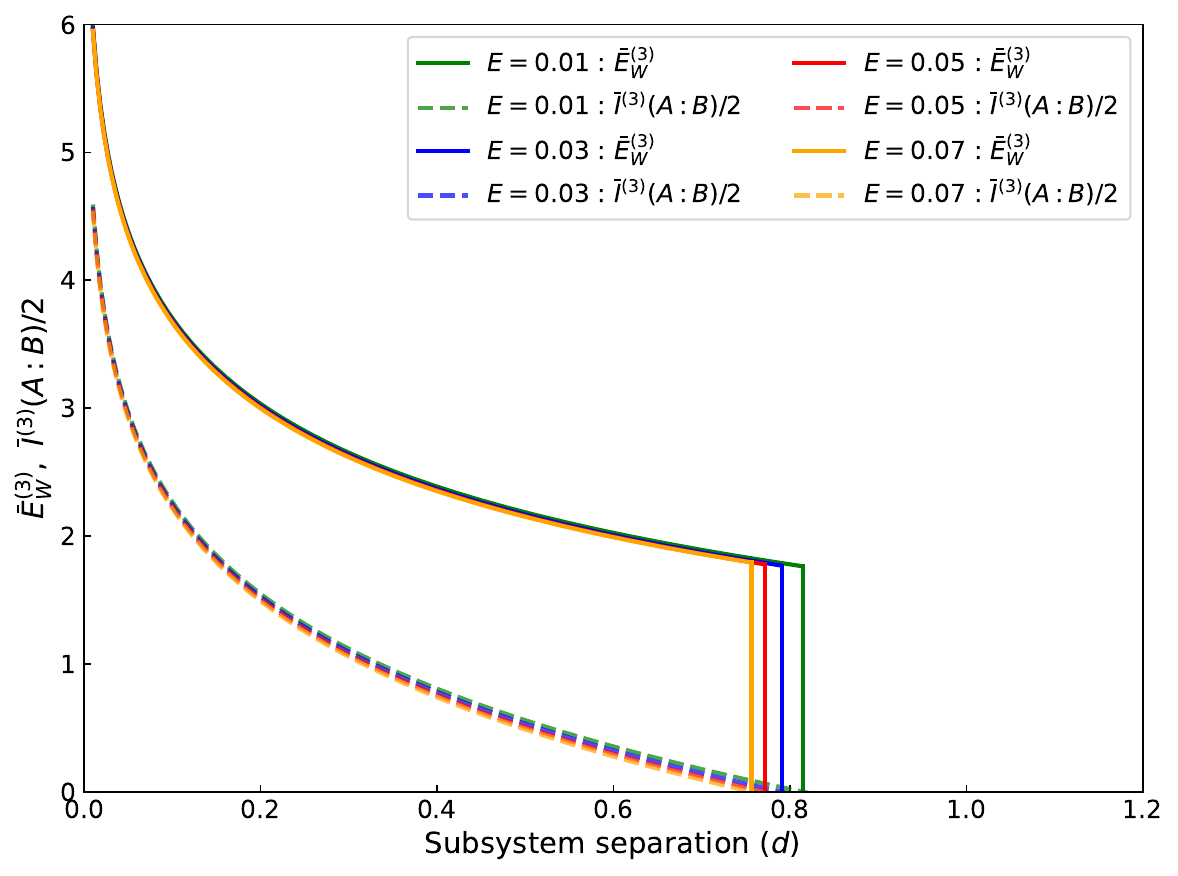}
         \caption{Plot for different values of the applied electric field.}
         \label{fig:EW3I diff E}
     \end{subfigure}
     ~~~
     \begin{subfigure}[t]{0.45\textwidth}
         \centering
         \includegraphics[width=\textwidth]{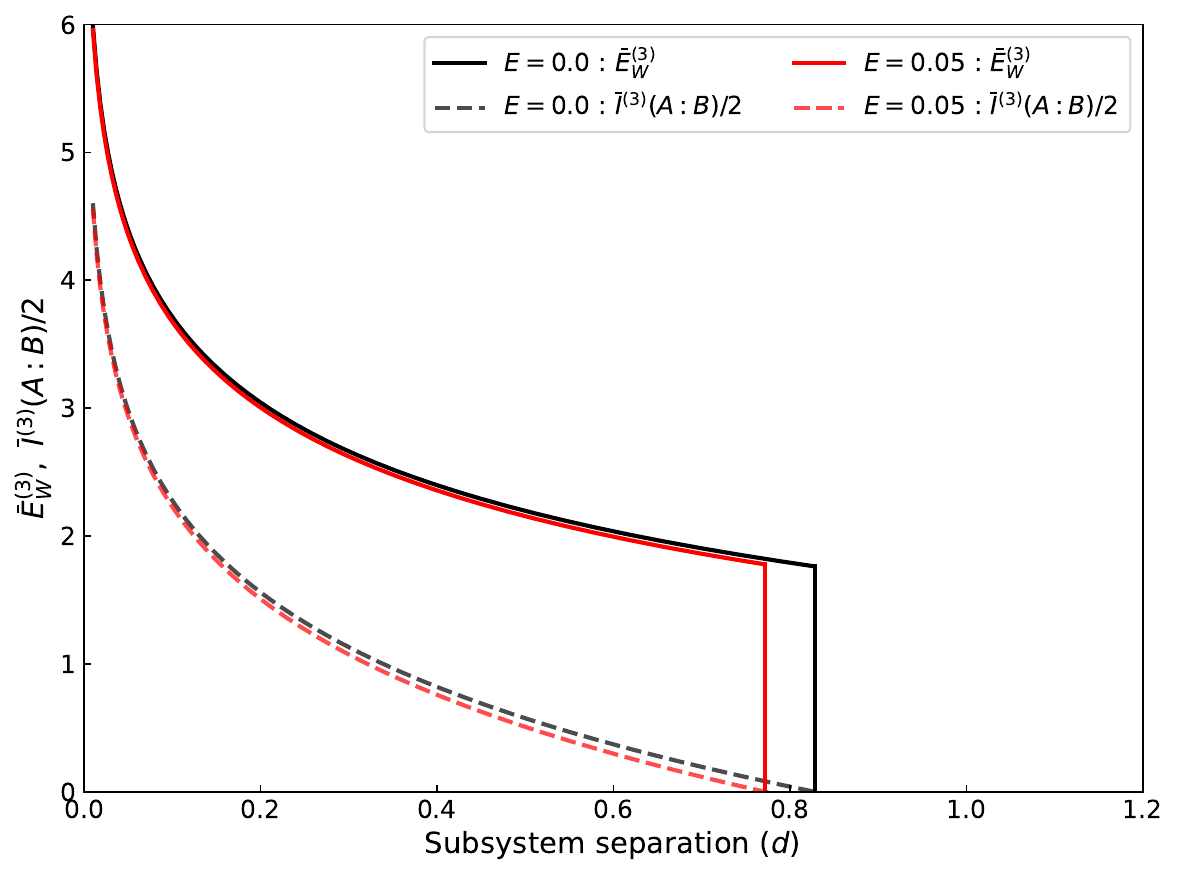}
         \caption{Comparison between the results of pure AdS$_3$.}
         \label{fig:EW3I comp}
     \end{subfigure}
        \caption{Variation of EWCS and HMI in case of $(2+1)$-dimensional OSM for two disjoint subsystems (with length $l=2$) with respect to their separation distance. In the left panel, we have shown the variation of EWCS and HMI with the separation distance between the two subsystems. The plots are done for four different values of the applied electric field ($E=0.01,~0.03,~0.05$ and $0.07$) in the super-gravity background. In the right panel of the figure, we have compared the results of EWCS and HMI for the applied electric field of $E=0.05$ with the pure AdS$_3$ result ($E=0$).}
        \label{fig:three graphs}
\end{figure}
With all these expressions in hand, we can proceed to find the critical subsystem separation distance $d_c$ where the mutual information vanishes for a fixed choice of subsystem length. It is discussed in \cite{Takayanagi:2017knl} that when the separation distance is less than the critical separation distance (that is, $d<d_c$), the entanglement wedge ($M_{AB}$) is in the connected phase. On the other hand, when $d>d_c$, the entanglement wedge is in a disconnected phase. Several interesting works in this direction can be found in \cite{Basu:2021awn,Sahraei:2021wqn,Chowdhury:2021idy,RoyChowdhury:2023iyr,Liu:2021rks,ChowdhuryRoy:2022dgo,Paul:2024lmd,BabaeiVelni:2023cge,Maulik:2022hty,Asadi:2022mvo,Yang:2023wuw,Yang:2025qkh,Jain:2022hxl,Gong:2020pse,Liu:2023rhd,Li:2023edb,Li:2021rff,Chen:2021bjt}. This discontinuity indeed represent a phase transition of EWCS from the connected phase to the disconnected phase. In order to obtain this critical separation distance $d_c$, one needs to solve the following equation
\begin{align}
    &\Big[4\log (l)-2\log (2l+d_c)-2\log (d_c)\Big]+\frac{E}{4}\Big[2l^2-(2l+d_c)^2-d_c^2\Big]\nonumber\\
    &-\frac{11E^2}{480}\Big[2l^4-(2l+d_c)^4-d_c^4\Big]=0~.
\end{align}
The above equation clearly suggests that the critical separation distance $d_c$ depends upon the subsystem length $l$ and the applied electric field $E$. This indeed means that for a given choice of $l$ and $E$, as long as the subsystem separation is smaller than the critical separation, that is $d<d_c$ the entanglement wedge is in the connected phase. Also, for subsystem separation greater than critical separation, that is $d>d_c$, the entanglement wedge for the total system is in a disconnected phase. \\
We have also graphically represented our results of EWCS and HMI with respect to the subsystem separation distance. In Fig.(\eqref{fig:EW3I diff E}), we have plotted the variation of EWCS and HMI with respect to subsystem separation distance in the case of $(2+1)$-dimensional OSM. In order to obtain this plot, we have set $l=2$ and chosen different values of the applied electric field ($E$). The curves in green, purple, red and orange are done for applied electric field values $E=0.01,~0.03,~0.05$ and $0.07$ respectively. Fig.(\eqref{fig:EW3I comp}) shows a comparison between the results of EWCS and HMI for a finite value of the applied electric field ($E=0.05$) with the corresponding results of pure AdS$_3$. The curve in black describes the pure AdS$_3$ result ($E=0$) and the curve in red is for the case of $E=0.05$. Both the Figures (\eqref{fig:EW3I diff E},\eqref{fig:EW3I comp}) tells that for increasing values of the applied electric field, the critical separation distance ($d_c$) at which the entanglement wedge becomes disconnected reduces. This implies that increasing the world volume electric field makes the entanglement wedge disconnected at smaller separation distances. From both the Figures, it is also evident that the well-known inequality between EWCS and HMI, that is $E_W\geq \frac{I}{2}$ \cite{Terhal:2002riz,Takayanagi:2017knl,Nguyen:2017yqw}, holds true.\\
Now we will also determine the change in the EWCS from the pure AdS$_3$ geometry due to the application of an electric field, which reads
\begin{align}
    \Delta\bar E^{(3)}_{W}&=\bar E^{(3)}_{W}-\bar E^{(AdS_3)}_{W}\nonumber\\&=-\frac{E}{4}l(d+l)+\frac{157E^2}{960}l(d+l)(d^2+2dl+2l^2)
\end{align}
\subsection{HMI and EWCS in (3+1)-dimensions}
In this subsection, we will compute the holographic mutual information and EWCS for a $(3+1)$-dimensional OSM. We will start with the derivation of HMI. Substituting the expression of HEE in eq.\eqref{HEE 3+1 final} into eq.\eqref{HMI main formula}, we get the following form of the HMI for $(3+1)$-dimensional OSM
\begin{align}
    I^{(2)}_{4}(A:B)=&\frac{L}{4G_4}\Bigg[4b_0^2\Big(-\frac{2}{l}+\frac{1}{2l+d}+\frac{1}{d}\Big)+\frac{E^2}{20b_0^2}\Big(2l^3-(2l+d)^3-d^3\Big)\nonumber\\&-\frac{E^4}{1600b_0^6}\Big(2l^7-(2l+d)^7-d^7\Big)\Bigg]
\end{align}
For further use and simplicity, we can define $\bar I^{(2)}_{4}(A:B)=\frac{4G_4}{L}I^{(2)}_{4}(A:B)$.\\
Now we will again proceed further to compute the entanglement wedge cross section for the $(3+1)$-dimensional OSM. We will follow the same approach to compute the EWCS as shown in the previous section. We will again consider two parallel boundary subsystems of length $l$ and separated by a distance $d$ such that $\frac{d}{l}\ll 1$. We need to calculate the vertical constant $x_1$ hypersurface with minimal area that splits the entanglement wedge $M_{AB}$ into two domains corresponding to $A$ and $B$. We will also assume the boundary coordinate $x_2$ is extended such that $0<x_2<L$. Also, we are dealing with static boundary field theories; therefore, the induced metric can be obtained by setting $dt=dx_1=0$. This gives
\begin{equation}
    dS^2_{ind}=\frac{1}{z^2}\Bigg[\frac{dz^2}{1-E^2 z^4}+dx_2^2\Bigg]~.
\end{equation}
The above induced metric can be used to obtain the expression of EWCS for $(3+1)$- dimensional OSM, which reads
\begin{align}
    E^{(4)}_W&=\frac{1}{4G_4}\int_{0}^{L}dx_2\int_{z_{*}(d)}^{z_{*}(2l+d)}\frac{dz}{z^2 \sqrt{1-E^2 z^4}}\nonumber\\
    &=\frac{L}{4G_4}\int_{z_{*}(d)}^{z_{*}(2l+d)}\frac{dz}{z^2 \sqrt{1-E^2 z^4}}~.
\end{align}
The above integral can be evaluated perturbatively, thus expanding the integrand considering $E$ to be a perturbation parameter and keeping terms up to $\mathcal{O}(E^4)$, we get
\begin{align}
    E^{(4)}_W&\approx\frac{L}{4G_4}\int_{z_{*}(d)}^{z_{*}(2l+d)}\frac{dz}{z^2}\Bigg[1+\frac{E^2}{2}z^4+\frac{3E^4}{8}z^8\Bigg]\nonumber\\
    &=\frac{L}{4G_4}\Bigg[\left\{\frac{1}{z_{*}(d)}-\frac{1}{z_{*}(2l+d)}\right\}+\frac{E^2}{6}\left\{z_{*}(2l+d)^3-z_{*}(d)^3\right\}\nonumber\\& +\frac{3E^4}{56}\left\{z_{*}(2l+d)^7-z_{*}(d)^7\right\}\Bigg]
\end{align}
Now we will move forward to express the EWCS in terms of the subsystem length $l$. In order to do the same, we will use eq.\eqref{zt 3+1}, which relates the turning point of the RT surface ($z_*$) to the subsystem length $l$. Hence, the above equation becomes
\begin{align}
    \bar E^{(4)}_W&=2b_0\Big(\frac{1}{d}-\frac{1}{2l+d}\Big)-\frac{E^2}{30b_0^3}\left\{l(3d^2+6dl+4l^2)\right\}\nonumber\\&+\frac{31 E^4}{44800 b_0^7}\left\{(2l+d)^7-d^7\right\}
\end{align}
where we have again defined $\bar E^{(4)}_W=\frac{4G_4}{L}E^{(4)}_W$.\\
\begin{figure}
     \centering
     \begin{subfigure}[t]{0.45\textwidth}
         \centering
         \includegraphics[width=\textwidth]{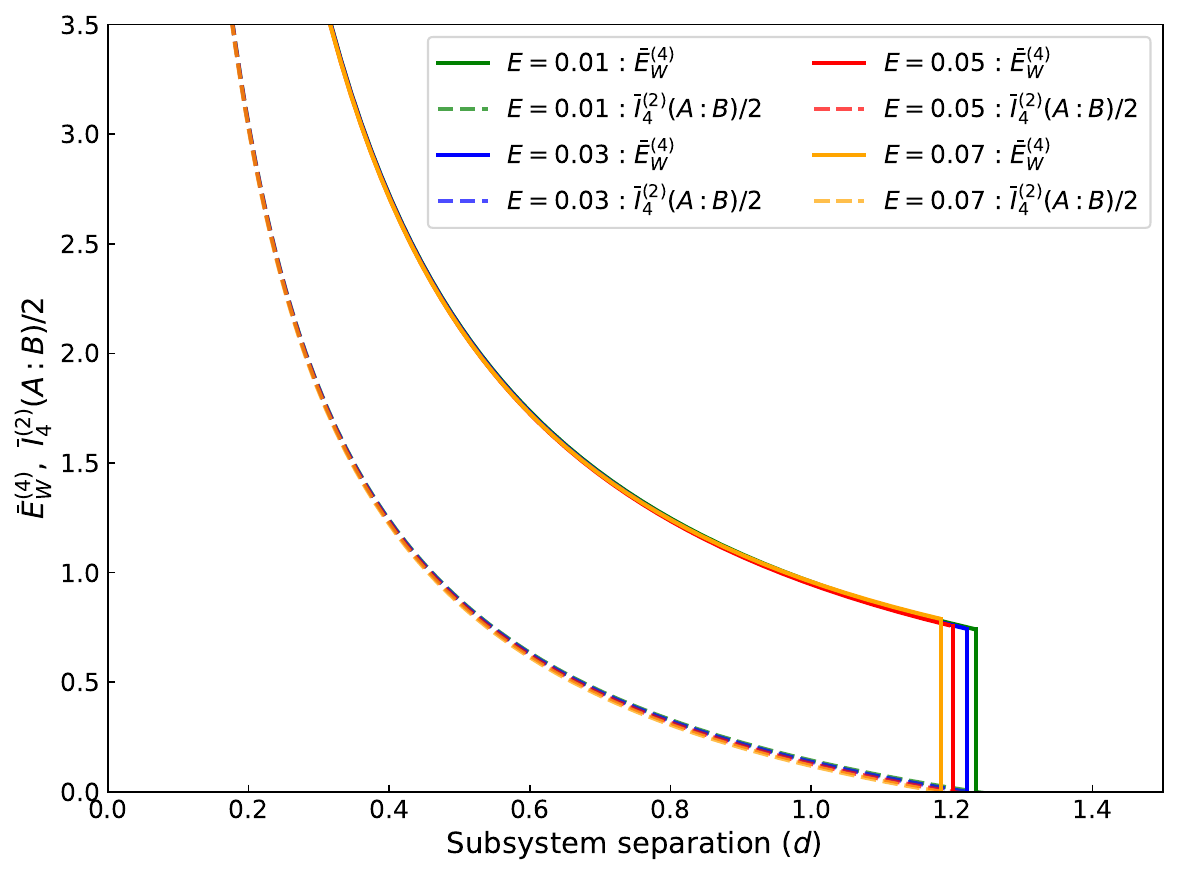}
         \caption{Plot for different values of the applied electric field.}
         \label{fig:EW4I diff E}
     \end{subfigure}
     ~~~
     \begin{subfigure}[t]{0.45\textwidth}
         \centering
         \includegraphics[width=\textwidth]{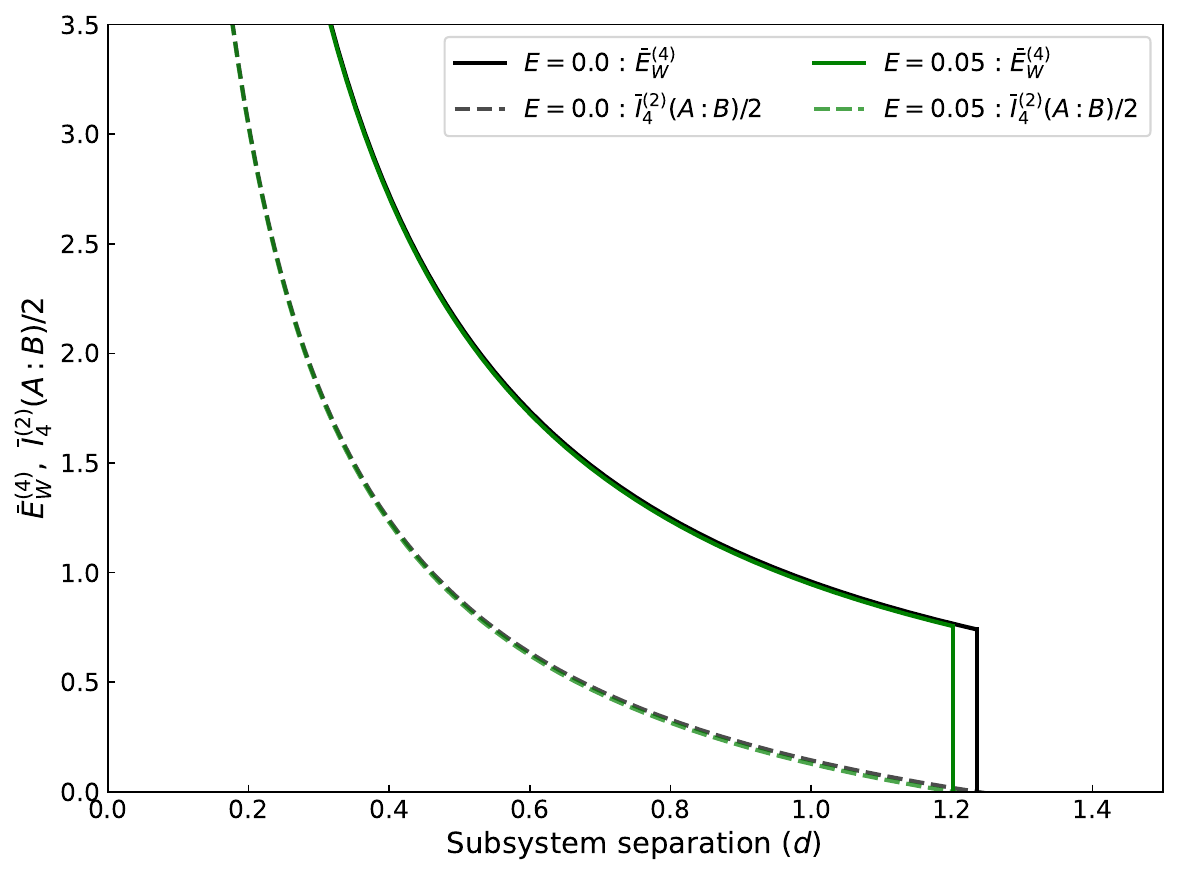}
         \caption{Comparison between the results of pure AdS$_4$.}
         \label{fig:EW4I comp}
     \end{subfigure}
        \caption{Variation of EWCS and HMI for two disjoint subsystems (each of length $l=2$) in a $(3+1)$-dimensional Open String Metric (OSM) background as a function of their separation distance. The left panel displays the dependence of EWCS and HMI on the separation for four different electric field values ($E=0.01$, $0.03$, $0.05$, and $0.07$) within the supergravity background. The right panel compares the results at $E=0.05$ with the corresponding pure AdS$_4$ case ($E=0$).}
        \label{fig:four graphs}
\end{figure}
With all these expressions in hand, we can proceed to find the critical subsystem separation distance $\tilde d_c$ where the mutual information vanishes for a fixed choice of subsystem length. Just like the previous subsection \eqref{subsec: 4.1}, we can evaluate the critical separation distance by setting HMI to be zero. Thus, in order to obtain this critical separation distance $\tilde d_c$, one needs to solve the following equation
\begin{align}
    &4b_0^2\Big(-\frac{2}{l}+\frac{1}{2l+\tilde d_c}+\frac{1}{\tilde d_c}\Big)+\frac{E^2}{20b_0^2}\Big(2l^3-(2l+\tilde d_c)^3-\tilde d_c^3\Big)\nonumber\\&-\frac{E^4}{1600b_0^6}\Big(2l^7-(2l+\tilde d_c)^7-\tilde d_c^7\Big)=0~.
\end{align}
The roots of the above equation clearly depend upon subsystem length ($l$) and applied electric field ($E$). Therefore, for a fixed choice of $l$ and $E$, as long as subsystem separation is less than critical separation, that is $d<\tilde d_c$, the entanglement wedge is in the connected phase; for $d>\tilde d_c$ the entanglement wedge becomes disconnected.\\
We have also presented our results for the EWCS and HMI graphically, as functions of the subsystem separation distance. In Fig.(\ref{fig:EW4I diff E}), we illustrate how both EWCS and HMI vary with the separation distance in the case of a $(3+1)$-dimensional OSM. To generate this plot, we fixed the parameter ($l=2$) and considered different values of the applied electric field ($E$). Specifically, the curves in green, purple, red, and orange correspond to electric field strengths ($E=0.01,~0.03,~0.05,~0.07$), respectively. Fig.(\ref{fig:EW3I comp}) compares the results of EWCS and HMI at a fixed electric field $E=0.05$ with those in pure AdS$_4$. The black curve represents the pure AdS$_4$ case ($E=0$), while the green curve shows the scenario at ($E=0.05$). Both figures demonstrate that increasing the applied electric field decreases the critical separation distance $\tilde d_c$ at which the entanglement wedge becomes disconnected. This indicates that a stronger electric field on the world volume causes the entanglement wedge to disconnect at smaller separations. Additionally, from both figures, it is clear that the well-known inequality $E_W \geq \frac{I}{2}$, relating EWCS and HMI, remains valid throughout the analysis.\\
Again, we can determine the change in EWCS for $(3+1)$-dimensional OSM from the pure AdS$_4$ case due to the application of a world volume electric field. The change is given by
\begin{align}
    \Delta\bar E^{(4)}_W=-\frac{E^2}{30b_0^3}\left\{l(3d^2+6dl+4l^2)\right\}+\frac{31 E^4}{44800 b_0^7}\left\{(2l+d)^7-d^7\right\}
\end{align}
\section{Holographic computation for entanglement negativity}\label{Sec:5}
In this section, we will briefly discuss about the calculation of the entanglement negativity of two parallel boundary subsystems in a holographic manner. The existing literature contains two different proposals for computing holographic entanglement negativity.\\
One of the proposals \cite{Kudler-Flam:2018qjo,Kusuki:2019zsp} suggests that entanglement negativity
($E_N$) is dual to EWCS backreacted by an extremal cosmic brane that terminates at the
boundary of the entanglement wedge. The motivation for this proposal comes from quantum error-correcting codes. Therefore, the logarithmic negativity is the same as the cross-sectional area of the entanglement wedge, along with some bulk contribution. However, the negativity for an arbitrary entangling surface is extremely hard to compute because of the backreaction from the cosmic brane. Although the calculation is quite simple for spherical entangling surfaces due to their symmetry. For a dual boundary CFT, the holographic entanglement negativity is given by
\begin{equation}
    E_N=\chi_{d}\frac{E_W}{4G_N}+E_{bulk}
\end{equation}
where $\chi_d$ is a dimension-dependent constant. It should be mentioned that for a spherical entangling surface, an explicit mathematical expression for $\chi_d$ can be found in \cite{Hung:2011nu}.\\
Another proposal \cite{Jain:2017uhe,Jain:2017xsu,Chaturvedi:2016rcn,Chaturvedi:2016rft} suggests that the entanglement negativity is certain combinations of co-dimension two RT surfaces. This type of combination is obtained from dual CFT correlators. One should mention that these two different proposals leads to the same result of entanglement negativity. Several interesting studies regarding the equivalence of these two proposals can be found in \cite{Rogerson:2022yim,Matsumura:2022ide,Bertini:2022fnr,Roik:2022gbb,Dong:2021oad,Afrasiar:2021hld,Bhattacharya:2021dnd,Hejazi:2021yhz}.\\
In the second approach, one can compute the entanglement negativity of two parallel boundary subsystems. At first, we will perform the analysis for two adjacent subsystems, then we will calculate entanglement negativity for two disjoint subsystems. We will also consider boundary subsystems with equal and unequal widths.
\subsection{Entanglement negativity in (2+1)-dimension}
This subsection is dedicated to the study of holographic entanglement negativity for a $(2+1)$-dimensional open string geometry. This system is holographic dual to quarks in a $(1+1)$-dimensional boundary quantum field theory.
\subsubsection{Adjoint subsystems}
The holographic entanglement negativity for two adjacent boundary subsystems of different widths ($l_1$ and $l_2$) is given by
\begin{equation}\label{EN adjoint}
    E^{(3)}_{N_{adj}}=\frac{3}{4}\Big[S^{(2)}_{HEE}(l_1)+S^{(2)}_{HEE}(l_2)-S^{(2)}_{HEE}(l_1+l_2)\Big]
\end{equation}
Using the expression of the HEE for $(2+1)$-dimensional OSM from eq.\eqref{HEE 2+1} in the above equation, one can write
\begin{align}
    E^{(3)}_{N_{adj}}&=\frac{3}{16G_3}\Bigg[2\log \Bigg(\frac{l_1l_2}{(l_1+l_2)\epsilon}\Bigg)+\frac{E}{4}\Big(l_1^2+l_2^2-(l_1+l_2)^2\Big)\nonumber\\
    &-\frac{11E^2}{480}\left\{l_1^4+l_2^4-(l_1+l_2)^4\right\}\Bigg]~.
\end{align}
For two adjoint subsystems of equal width $l_1=l_2=l$. Hence the above equation simplifies to the following form
\begin{equation}
    \bar E^{(3)}_{N_{adj}}=\frac{3}{4}\Bigg[2\ln \Big(\frac{l}{2\epsilon}\Big)-\frac{E}{2}l^2+\frac{77E^2}{240}l^4\Bigg]~.
\end{equation}
where $\bar E^{(3)}_{N_{adj}}=4G_3 E^{(3)}_{N_{adj}}$.
\begin{figure}
     \centering
     \begin{subfigure}[t]{0.45\textwidth}
         \centering
         \includegraphics[width=\textwidth]{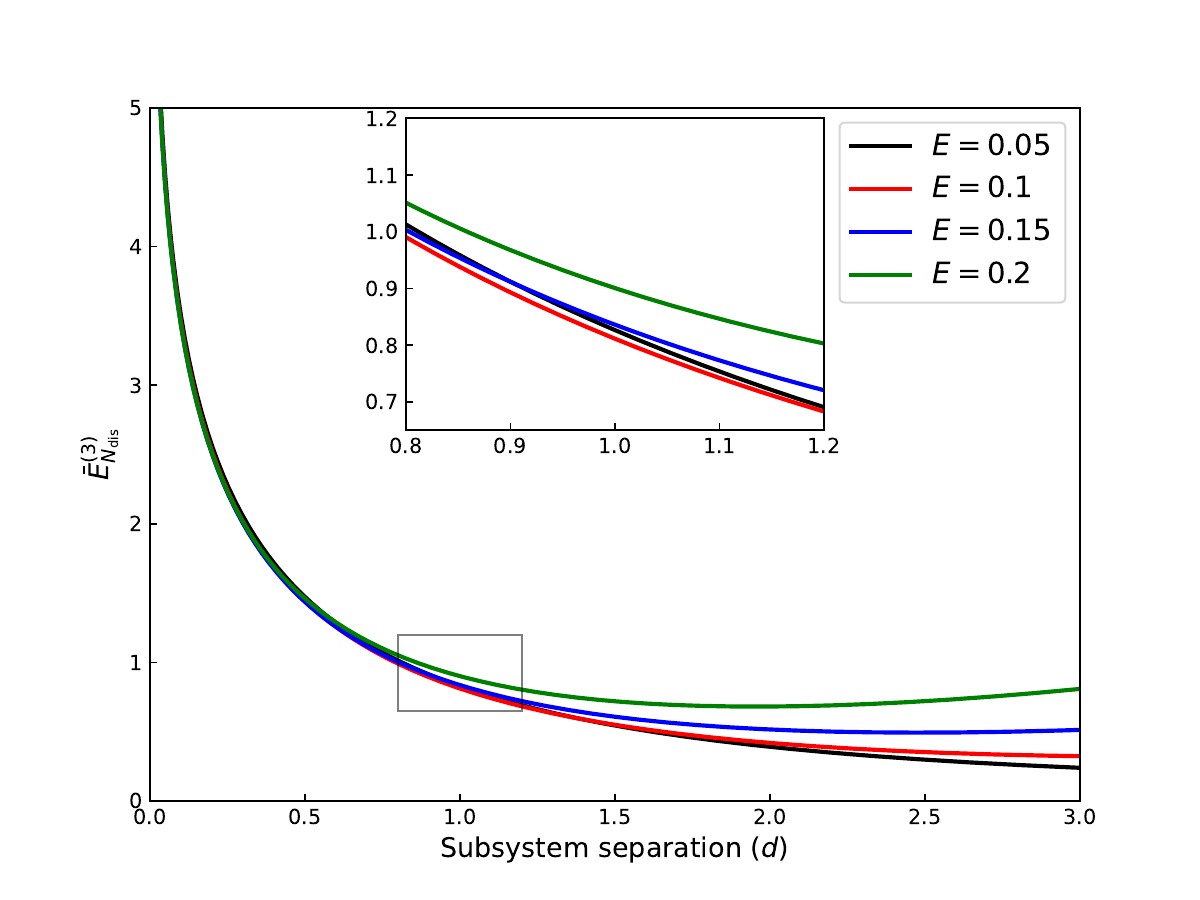}
         \caption{Plot for different values of the applied electric field.}
         \label{fig:EN3 diff E}
     \end{subfigure}
     ~~
     \begin{subfigure}[t]{0.45\textwidth}
         \centering
         \includegraphics[width=\textwidth]{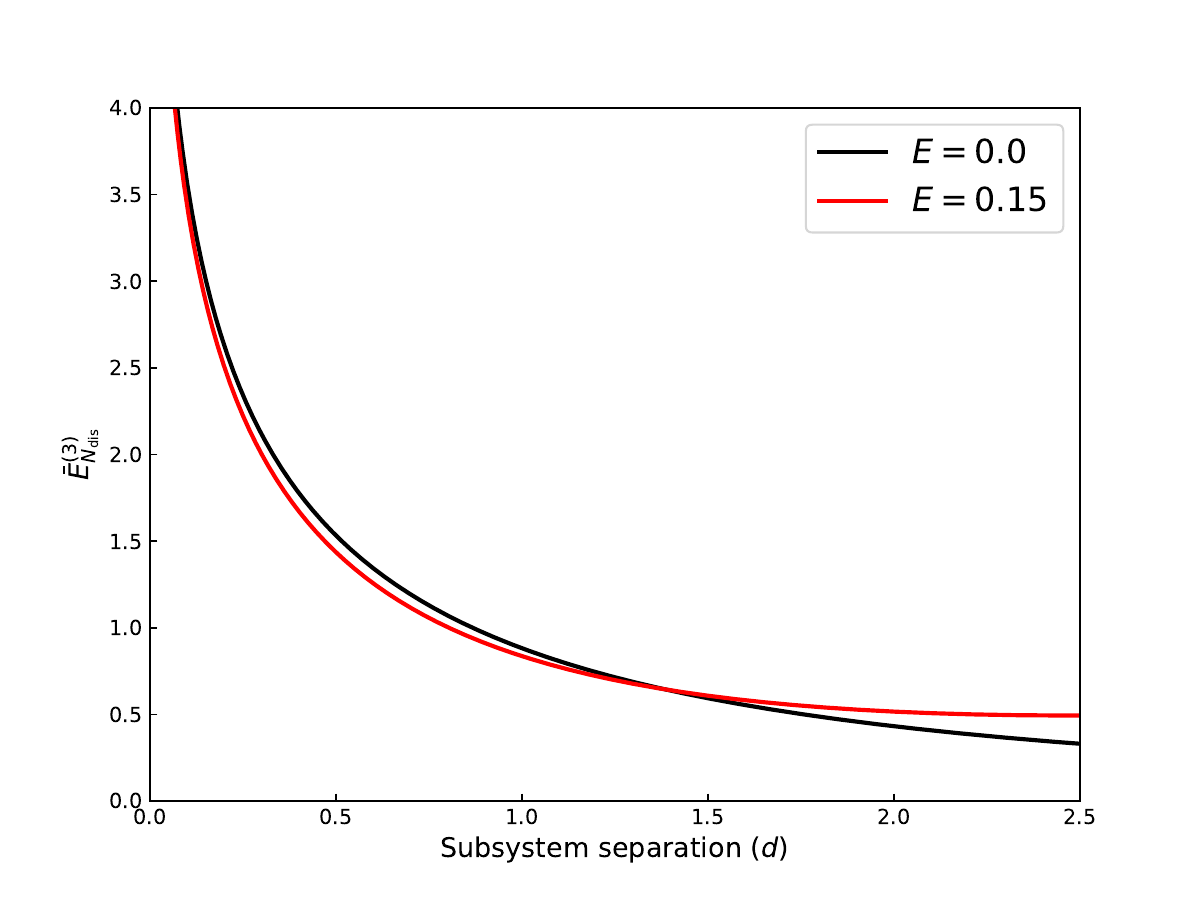}
         \caption{Comparison between the results of pure AdS$_3$.}
         \label{fig:EN3 comp}
     \end{subfigure}
        \caption{Left panel shows the variation of the holographic entanglement negativity \(\bar{E}^{(3)}_{N_{dis}}\) for two disjoint parallel boundary subsystems of equal width (\(l=2\)) as a function of their separation \(d\), for different electric field strengths (\(E=0.05,~0.1,~0.15,\) and \(0.2\)). Right panel shows the comparison between the pure AdS\(_3\) background and the OSM with an applied electric field \(E=0.15\).}
        \label{fig:EN3}
\end{figure}
Hence, for two adjacent subsystems of equal width ($l$) the change in entanglement negativity from the pure AdS$_3$ scenario is given by
\begin{equation}
    \Delta \bar E^{(3)}_{N_{adj}}=\bar E^{(3)}_{N_{adj}}-\bar E^{(2)}_{AdS_3}=\frac{3El^2}{32G_3}\Bigg[\frac{77El^2}{120}-1\Bigg]~.
\end{equation}
\subsubsection{Disjoint subsystems}
For two disjoint parallel boundary subsystems of widths $l_1$ and $l_2$ and separated by a distance $d$, the holographic entanglement negativity is given by
\begin{equation}\label{EN disjoint}
    E^{(2)}_{N_{dis}}=\frac{3}{4}\Big[S^{(2)}_{HEE}(l_1+d)+S^{(2)}_{HEE}(l_2+d)-S^{(2)}_{HEE}(l_1+l_2+d)-S^{(2)}_{HEE}(d)\Big]
\end{equation}
After putting the explicit expressions for $S^{(2)}_{HEE}(l_1+d)$,$S^{(2)}_{HEE}(l_2+d)$,$S^{(2)}_{HEE}(l_1+l_2+d)$ and $S^{(2)}_{HEE}(d)$ in the above equation using eq.\eqref{HEE 2+1}, we obtain
\begin{align}
    E^{(2)}_{N_{dis}}&=\frac{3}{16G_3}\Bigg[2\left\{\ln \Big(\frac{l_1+d}{\epsilon}\Big)+\ln\Big(\frac{l_2+d}{\epsilon}\Big)-\ln\Big(\frac{l_1+l_2+d}{\epsilon}\Big)-\ln\Big(\frac{d}{\epsilon}\Big)\right\}\nonumber\\
    &+\frac{E}{4}\left\{(l_1+d)^2+(l_2+d)^2-(l_1+l_2+d)^2-d^2\right\}\nonumber\\
    &-\frac{11E^2}{480}\left\{(l_1+d)^4+(l_2+d)^4-(l_1+l_2+d)^4-d^4\right\}\Bigg]~.
\end{align}
Now when the two disjoint boundary subsystems have equal width, we can choose $l_1=l_2=l$. This assumption leads to the following expression of the entanglement negativity
\begin{align}
    \bar E^{(3)}_{N_{dis}}&=\frac{3}{4}\Bigg[2\left\{2\ln (l+d)-\ln(2l+d)-\ln(d)\right\}\nonumber\\
    &+\frac{E}{4}\left\{2(l+d)^2-(2l+d)^2-d^2\right\}
    -\frac{11E^2}{480}\left\{2(l+d)^4-(2l+d)^4-d^4\right\}\Bigg]~.
\end{align}
where we have defined $\bar E^{(3)}_{N_{dis}}=4G_3 E^{(3)}_{N_{dis}}$.\\
Now we will move forward to show the graphical representation of HNI for two disjoint subsystems of equal width ($l$) with respect to subsystem separation ($d$) in the case of ($2+1$)-dimensional OSM. In Fig.(\ref{fig:EN3 diff E}), we have shown the variation of $\bar E^{(3)}_{N_{dis}}$ with respect to subsystem separation $d$. The graphs in black, red, blue and green correspond to $E=0.05,~0.1,~0.15$ and $0.2$ respectively. The right panel (Fig.\eqref{fig:EN3 comp}) shows the comparison between the results of HNI for pure AdS$_3$ geometry and OSM with $E=0.15$. The curves in black and red are for $E=0$ and $E=0.15$ respectively.  It should be mentioned that as our open string metric results from applying small values of electric field in the supergravity background, the HNI do not vanish at a critical separation distance. Consequently, while both the HMI and EWCS disappear at a finite critical separation, the HNI persists; therefore, the HNI serves as a more reliable indicator of entanglement in mixed states.

\noindent 
Now, for disjoint subsystems, the change in the entanglement negativity from the pure AdS$_3$ geometry can also be measured, which is given by
\begin{align}
    \Delta \bar E^{(3)}_{N_{dis}}&=\bar E^{(3)}_{N_{dis}}-\bar E^{(3)}_{AdS_{3}}\nonumber\\&=\frac{3E}{64G_3}\Bigg[\left\{2(l+d)^2-(2l+d)^2-d^2\right\}-\frac{11E}{120}\left\{2(l+d)^4-(2l+d)^4-d^4\right\}\Bigg]
\end{align}
\subsection{Entanglement negativity in (3+1)-dimension}
In this subsection, we will study the holographic entanglement negativity for a $(3+1)$-dimensional open string geometry. From the gauge/gravity framework, one can say that this system is holographically dual to quarks in a $(2+1)$-dimensional boundary quantum field theory.
\subsubsection{Adjoint subsystems}
For two parallel strip-like boundary subsystem of width $l_1$ and $l_2$, the entanglement negativity for $(3+1)$-dimensional OSM can be found by using the result of eq.\eqref{HEE 3+1 final} in eq.\eqref{EN adjoint}, which reads
\begin{align}
    E^{(4)}_{N_{adj}}=&\frac{3L}{16 G_4}\Bigg[\frac{2}{\epsilon}+4b_0^2\left\{\frac{1}{l_1+l_2}-\frac{1}{l_1}-\frac{1}{l_2}\right\}+\frac{E^2}{20b_0^2}\left\{l_1^3+l_2^3-(l_1+l_2)^3\right\}\nonumber\\& +\frac{E^4}{1600 b_0^6}\left\{(l_1+l_2)^7-l_1^7-l_2^7\right\}\Bigg]
\end{align}
Now we can also evaluate the expression for entanglement negativity for two adjacent parallel boundary subsystems of equal width (say $l_1=l_2=l$). The entanglement negativity in this scenario is given by
\begin{equation}
    \bar E^{(4)}_{N_{adj}}=\frac{3}{4}\Bigg[\frac{2}{\epsilon}-\frac{6b_0^2}{l}-\frac{3E^2}{10b_0^2}l^3+\frac{63E^4}{800}l^7\Bigg]~.
\end{equation}
where we have agian defined $\bar E^{(4)}_{N_{adj}}=\frac{4G_4}{L}$.\\
Thus, the change in entanglement negativity from the unperturbed pure AdS$_4$ case is given by
\begin{align}
    \Delta \bar E^{(4)}_{N_{adj}}&=\bar E^{(4)}_{N_{adj}}-\bar E^{(4)}_{AdS_{4}}\nonumber\\
    &=\frac{3L}{16 G_4}\Bigg[-\frac{3E^2}{10b_0^2}l^3+\frac{63E^4}{800}l^7\Bigg]
\end{align}
\begin{figure}
     \centering
     \begin{subfigure}[t]{0.45\textwidth}
         \centering
         \includegraphics[width=\textwidth]{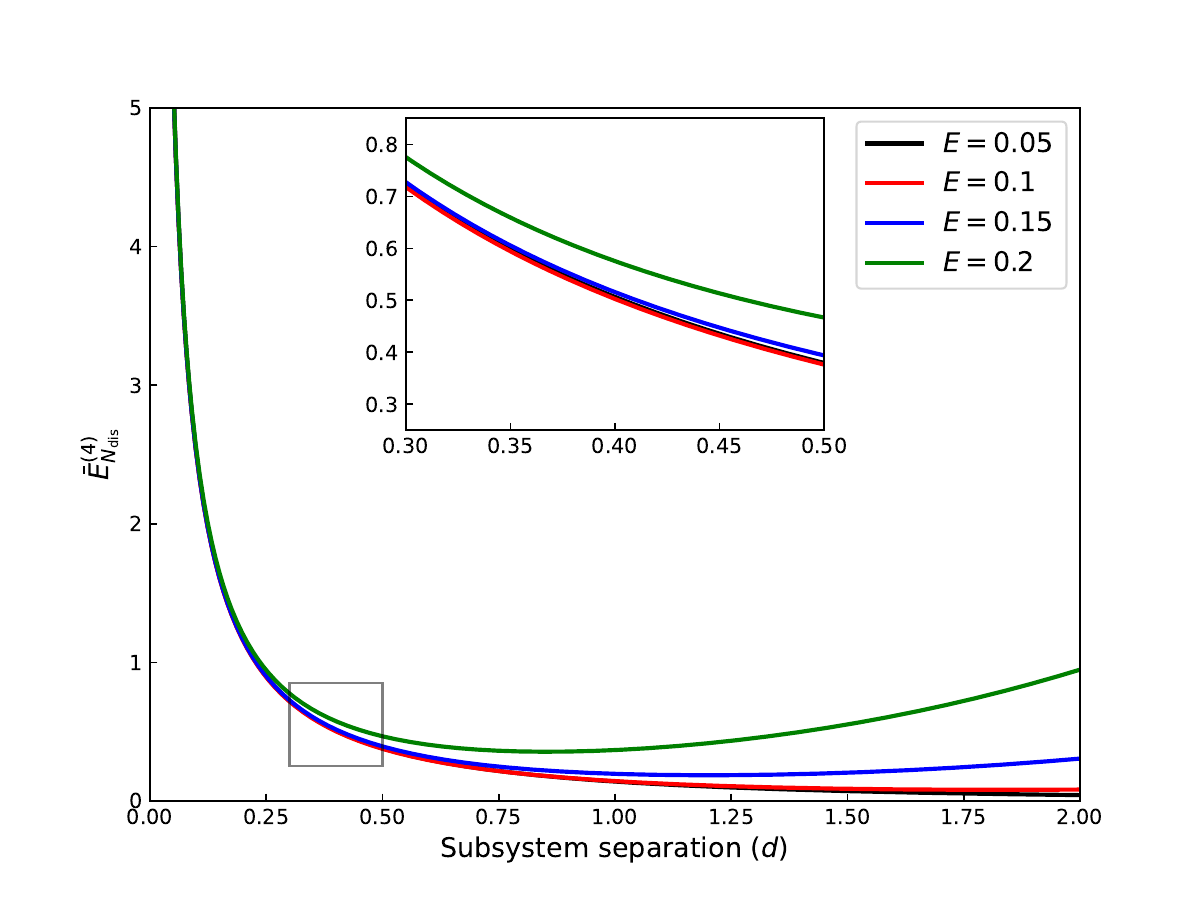}
         \caption{Plot for different values of the applied electric field.}
         \label{fig:EN4 diff E}
     \end{subfigure}
     ~~
     \begin{subfigure}[t]{0.45\textwidth}
         \centering
         \includegraphics[width=\textwidth]{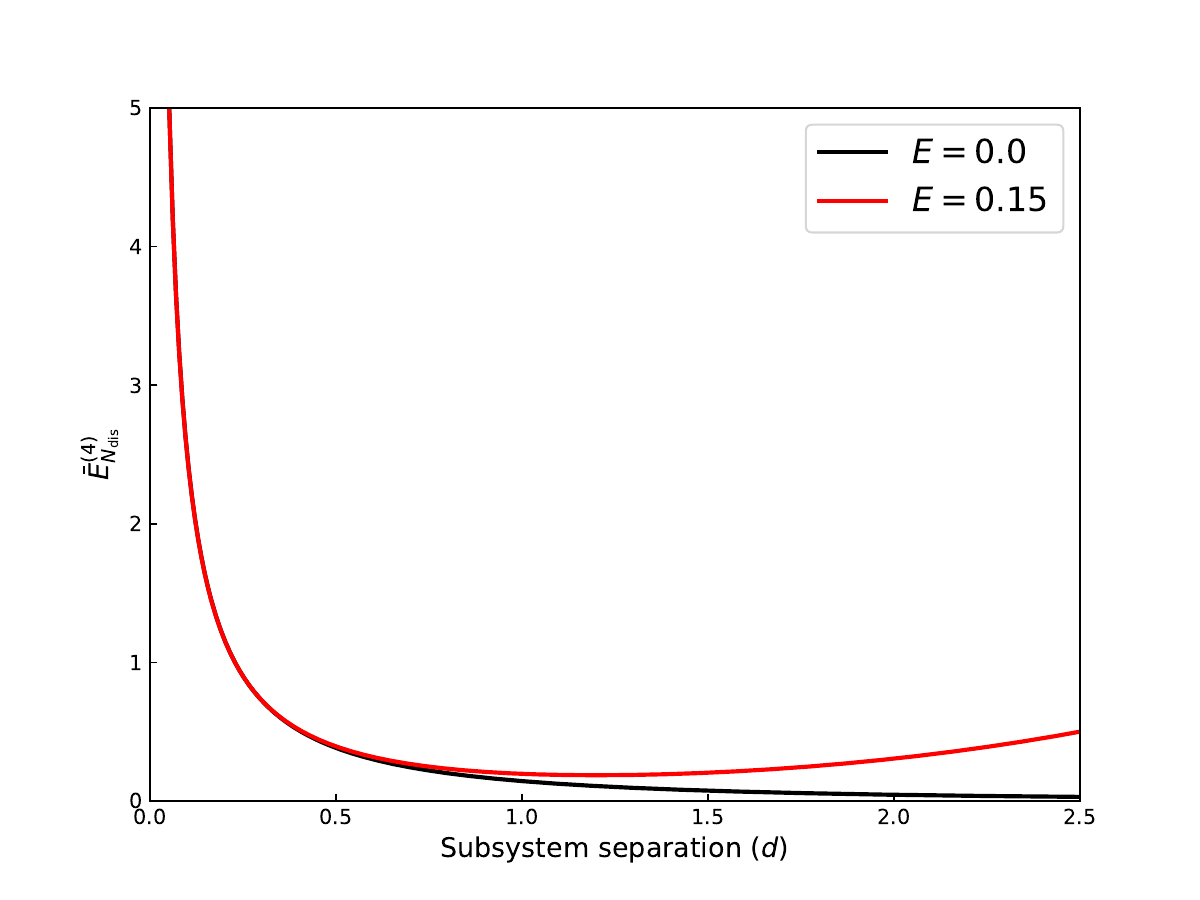}
         \caption{Comparison between the results of pure AdS$_4$.}
         \label{fig:EN4 comp}
     \end{subfigure}
        \caption{The left panel shows the variation of holographic entanglement negativity $\bar E^{(4)}_{N_{dis}}$ of two disjoint parallel boundary subsystems of equal width ($l=2$), with respect to the subsystem separation ($d$) for different values of the applied electric field ($E=0.05,~0.1,~0.15$ and $0.2$). In the right panel, we compare the results of pure AdS$_4$ with OSM having $E=0.15$ as the applied electric field value.}
        \label{fig:EN4}
\end{figure}
\subsubsection{Disjoint subsystems}
We have already seen in the subsection that for two disjoint parallel boundary subsystems of widths $l_1$ and $l_2$ and separated by a distance $d$, the holographic entanglement negativity is given by eq.\eqref{EN disjoint}. Hence, in order to evaluate the holographic entanglement negativity in this scenario, we will substitute the expression of HEE for $(3+1)$-dimensional OSM into eq.\eqref{EN disjoint}. This gives
\begin{align}
    E^{(4)}_{N_{dis}}=&\frac{3L}{16 G_4}\Bigg[4b_0^2\left\{\frac{1}{l_1+l_2+d}+\frac{1}{d}-\frac{1}{l_1+d}-\frac{1}{l_2+d}\right\}\nonumber\\&+\frac{E^2}{20b_0^2}\left\{(l_1+d)^3+(l_2+d)^3-(l_1+l_2+d)^3-d^3\right\}\nonumber\\&+\frac{E^4}{1600 b_0^6}\left\{(l_1+l_2+d)^7+d^7-(l_1+d)^7-(l_2+d)^7\right\}\Bigg]~.
\end{align}
In case of two parallel disjoint boundary subsystems of equal width ($l_1=l_2=l$), the above equation simplifies to the following 
\begin{align}
    \bar E^{(4)}_{N_{dis}}=&\frac{3}{4}\Bigg[4b_0^2\left\{\frac{1}{2l+d}+\frac{1}{d}-\frac{2}{l+d}\right\}+\frac{E^2}{20b_0^2}\left\{2(l+d)^3-(2l+d)^3-d^3\right\}\nonumber\\&+\frac{E^4}{1600 b_0^6}\left\{(2l+d)^7+d^7-2(l+d)^7\right\}\Bigg]~
\end{align}
where $\bar E^{(4)}_{N_{dis}}=\frac{4G_4}{L} E^{(4)}_{N_{dis}}$.\\
We again proceed to illustrate the graphical representation of HNI for two disjoint subsystems of equal width ($l$), considering the effect of subsystem separation ($d$) within the framework of ($3+1$)-dimensional OSM. In Fig. (\ref{fig:EN4 diff E}), we display the variation of $\bar E^{(3)}_{N_{dis}}$ as a function of the separation distance $d$. The curves in black, red, blue, and green correspond to electric field values $E=0.05$, $0.1$, $0.15$, and $0.2$, respectively. In Fig. (\ref{fig:EN4 comp}), we compare the results of HNI for pure AdS$_4$ geometry and OSM with $E=0.15$. The curves in black and red are for $E=0$ and $E=0.15$ respectively. It is noteworthy that, since our open string metric results are obtained under small electric field values in the supergravity background, the HNI does not vanish at a critical separation distance. Unlike the HMI and EWCS, which disappear at finite critical separations, the HNI remains non-zero; thus, the HNI serves as a more reliable measure of entanglement in mixed states.
Again, for the disjoint scenario, the change in entanglement negativity from the results of pure AdS$_4$ geometry can be calculated, which reads
\begin{align}
    \Delta \bar E^{(4)}_{N_{adj}}&=\bar E^{(4)}_{N_{adj}}-\bar E^{(4)}_{AdS_{4}}\nonumber\\
    &=\frac{3}{4}\Bigg[\frac{E^2}{20b_0^2}\left\{2(l+d)^3-(2l+d)^3-d^3\right\}+\frac{E^4}{1600 b_0^6}\left\{(2l+d)^7+d^7-2(l+d)^7\right\}\Bigg]
\end{align}
\section{Conclusion}\label{Sec:6}
Now we will summarize our findings. In this paper, we have studied various mixed state information theoretic quantities for open string geometries in $(2+1)$ and $(3+1)$-dimensions. The supergravity background is generated by a stack of $D$ branes. It is well known that the gauge theory living in the world volume of these $D$-branes (color branes) has only the adjoint degrees of freedom. On the other hand, adding flavor degrees of freedom in the fundamental representation of the gauge group requires adding something called the flavor branes in the gravity dual description. To avoid the backreaction of these flavor branes on the supergravity background, one needs to have the flavor branes in the probe limit such that the number of flavor branes are much smaller than the color branes. Therefore, studying various mixed-state information-theoretic quantities for open string geometries is equivalent to studying the mixed-state information-theoretic measures in the flavor sector on the dual gauge-theory side. In the first part of our paper, we have briefly reviewed how the open string geometry arises while studying various scalar, vector and spinor fluctuations on the flavor branes. It was shown that for a specific choice of the world volume gauge field, a horizon structure emerges in the open string geometry. Considering the AdS$_3$ spacetime as background, we have briefly derived the OSM in $(2+1)$-dimensions. Then cosnidering strip like boundary subsystem, we have computed the HEE for $(2+1)$ and $(3+1)$-dimensional OSMs. The change in the HEE from the pure AdS geometry is also computed and has been graphically represented as contour plots. The contour plots show the change in HEE expressions with respect to two independent parameters, namely subsystem length ($l$) and the applied electric field ($E$). We then calculated the holographic mutual information (HMI) and the EWCS for the same set of geometries. To do so, we have again considered two parallel strip-like boundary subsystems of equal width and separated by some finite distance. The graphical representation of these quantities shows that the entanglement wedge becomes disconnected at a critical subsystem separation distance where the mutual information vanishes. Our plots clearly show that the well-known inequality between the HMI and EWCS, that is $E_{W}\geq \frac{I}{2}$, holds true for OSMs. We have also compared the results of HMI and EWCS between the results of OSMs and the pure AdS case through plots. These plots show that the application of an electric field actually reduces the critical separation where the entanglement wedge becomes disconnected.  We have finally computed the holographic entanglement negativity again for $(2+1)$ and $(3+1)$-dimensional OSMs. We would like to mention that the calculations are done for both adjacent and disjoint subsystems. For both cases, we have considered parallel strip-like subsystems of both equal or unequal lengths. Our calculations reveal that the holographic entanglement negativity for the adjacent scenario (for subsystems with equal and unequal lengths) has a UV divergent pice. Similarly the holographic entanglement negativity for disjoint subsystems of different lengths has a UV divergent part.  On the other hand, the holographic entanglement negativity is free from any UV divergence for the disjoint scenario with equal subsystem lengths. We have also mentioned that as the parent metrics in the OSMs are chosen as pure AdS metrics, the entanglement negativity does not vanish at a finite critical separation. The graphical representation of entanglement negativity for the disjoint scenario is done for different values of the applied electric field. We have also shown a graphical comparison between the results of the pure AdS scenario and OSMs. In this sense, beyond a certain critical separation where the EWCS becomes disconnected, the entanglement negativity is still non-zero. This indeed implies that entanglement negativity is a better measure for mixed-state entanglement in the flavor sector of the dual boundary gauge theory. As a future direction of this work, one can explore the multipartite entanglement in these OSMs. Also the study of chaos in these spacetime backgrounds is still unexplored in the existing literature; hence, studying chaotic properties in these backgrounds may help us to study chaotic dynamics in the flavor sector of the boundary field theory.
\appendix
\renewcommand{\theequation}{A\arabic{equation}}
\setcounter{equation}{0}
\section{Computation for the diagonal OSM components}
In this appendix, we will calculate the diagonal OSM components $\tilde S_{xx}$, $\tilde S_{zz}$ and $\tilde S_{\tau\tau}$ in details. In order to do the same, we will use eq.\eqref{S diag}.\\
Comparing the $xx$ compnents on the both side of eq.\eqref{S diag}, we get
\begin{align}
    \tilde S_{xx}&=\frac{1}{z^2}-E^2z^2+a^{\prime 2}_x z^2\nonumber\\
    &=\frac{1}{z^2}-E^2z^2+\frac{j^2}{1-j^2 z^2}(1-E^2 z^4)~.
\end{align}
Now, from the reality condition of $a^{\prime}_x$ in eq.\eqref{aprime x}, we can use $j=\sqrt{E}$ in the above equation. This gives
\begin{equation}
    \tilde S_{xx}=\frac{1}{z^2}+E=\frac{1}{z^2}+\frac{1}{z_h^2}~.
\end{equation}
In the above equation, we have used the fact that the horizon of the OSM is situated at $z_h=\frac{1}{\sqrt{E}}$. If we consider the $zz$ component on both sides of eq.\eqref{S diag}, this gives
\begin{align}
    \tilde S_{zz}&=S_{tt}f^{\prime}(z)^2-S_{tz}f^{\prime}(z)+S_{zz}\nonumber\\
    &=\frac{-E^2 j^2 z^4}{(1-j^2 z^2)}+\frac{2E^2 jz^5 a^{\prime}_x}{\sqrt{(1-E^2 z^4)(1-j^2z^2)}}+\frac{1}{z^2}+a^{\prime 2}_x z^2
\end{align}
Now using the value of $a^{\prime}_x$ from eq.\eqref{aprime x}, we finaly get
\begin{align}
    \tilde S_{zz}&=\frac{E^2j^2z^4}{1-j^2z^2}+\frac{1}{z^2}+j^2 \Big(\frac{1-E^2 z^4}{1-j^2z^2}\Big)\nonumber\\&=\frac{j^2}{1-j^2z^2}+\frac{1}{z^2}=\frac{1}{z^2(1-\frac{z^2}{z_h^2})}~.
\end{align}
In the last line, we have used the fact that $j=\sqrt{E}$ and $z_h=\frac{1}{\sqrt{E}}$.
Finally considering the $\tau\tau$ component we get
\begin{equation}
    \tilde S_{\tau\tau}=S_{tt}=-\frac{1}{z^2}(1-E^2 z^4)=-\frac{1}{z^2}\Big(1-\frac{z^4}{z_h^4}\Big)~.
\end{equation}
Therefore, collecting all the metric elements, we can write the total OSM in $(2+1)$-dimensions as follows
\begin{align}
    ds^2_{osm}&=\tilde S_{\tau\tau}d\tau^2 +\tilde S_{xx}dx^2+\tilde S_{zz}dz^2\nonumber\\
    &=-\frac{1}{z^2}\Big(1-\frac{z^4}{z_h^4}\Big)d\tau^2+\Big(\frac{1}{z^2}+\frac{1}{z_h^2}\Big)dx^2+\frac{dz^2}{z^2\Big(1-\frac{z^2}{z_h^2}\Big)}~.
\end{align}
\section*{Acknowledgment}
SP thanks SNBNCBS for the Senior Research Fellowship. ARC would like to thank SNBNCBS for the fellowship.
\subsection*{Data Availability Statement}
This article has no associated data, or the data will not be deposited.
\subsection*{Code Availability Statement} 
This article has no associated code, or the code will not be deposited.

\bibliographystyle{JHEP}
\bibliography{ref-osm}
\end{document}